\UseRawInputEncoding
\documentclass[aps,10pt,pre,twocolumn,amsmath,amssymb,amsfonts]{revtex4-2}

\usepackage{graphicx}
\usepackage{hyperref}

\def\Eq#1{Eq.(\ref{#1})}
\def\Eqs#1{Eqs.(\ref{#1})}
\def\Fig#1{Fig.~\ref{#1}}

\def\nm{\nonumber\\}
\def\ds{\displaystyle}
\def\>{\rangle}
\def\<{\langle}

\def\half{\frac{1}{2}}
\def\ch{\text{ch}}
\def\tot{\text{tot}}
\def\adg{a^\dagger}
\def\Adg{A^\dagger}
\def\Ldg{L^\dagger}
\def\bdg{b^\dagger}
\def\cdg{c^\dagger}
\def\dg{\dagger}
\def\lam{\lambda}
\def\w{\omega}
\def\del{\delta}
\def\gam{\gamma}

\def\Gam{\Gamma}
\def\al{\alpha}
\def\bt{\beta}
\def\sig{\sigma}
\def\kap{\kappa}
\def\eps{\epsilon}

\def\tr{\text{tr}}
\def\si{\text{sk}}
\def\s{\text{s}}
\def\src{\text{src}}
\def\rad{\text{r}}
\def\st{\text{site}}
\def\diss{\text{diss}}
\def\ph{\text{ph}}
\def\ext{\text{ext}}
\def\th{\text{t$\!$h}}
\def\cmi{\text{cm}^{-1}}
\def\psi{\text{ps}^{-1}}

\def\smn{\sideset{}{'}\sum_{\mu<\nu}}
\def\siL{\sum_{x=1}^\ell}
\def\siLm{\sum_{x=1}^{\ell-1}}
\def\smL{\sum_{\mu=1}^\ell}
\def\ssL{\sum_{\sig=1}^\ell}
\def\mn{{\mu\nu}}
\def\ms{{\mu\sig}}
\def\sn{{\sig\nu}}
\def\sk{{\sig\kappa}}
\def\mk{{\mu\kappa}}
\def\kn{{\kappa\nu}}
\def\nb{\bar{n}}
\def\d{\partial}

\begin{document}

\title{Excitation relaxation in molecular chain and energy transfer at steady state}

\author{B.~A.~Tay}
\email{BuangAnn.Tay@nottingham.edu.my}
\affiliation{Department of Foundation in Engineering, Faculty of Science and Engineering, University of Nottingham Malaysia, Jalan Broga, 43500 Semenyih, Selangor, Malaysia}

\date{\today}

\begin{abstract}
We consider the reduced dynamics of a molecular chain weakly coupled to a phonon bath.  With a small and constant inhomogeneity in the coupling, the excitation relaxation rates are obtained in closed form. They are dominated by transitions between exciton modes lying next to each other in the energy spectrum. The rates are quadratic in the number of sites in a long chain. Consequently, the evolution of site occupation numbers exhibits longer coherence lifetime for short chains only. When external source and sink are added, the rate equations of exciton occupation numbers are similar to those obtained earlier by Fr\"{o}hlich to explain energy storage and energy transfer in biological systems. There is a clear separation of time scale into a faster one pertaining to internal influence of the chain and phonon bath, and a slower one determined by external influence, such as the pumping rate of the source, the absorption rate of the sink and the rate of radiation loss. The energy transfer efficiency at steady state depends strongly on these external parameters, and is robust against a change in the internal parameters, such as temperature and inhomogeneity. Excitations are predicted to concentrate to the lowest energy mode when the source power is sufficiently high. In the site basis, this implies that when sustained by a high power source, a sink positioned at the center of the chain is more efficient in trapping energy than a sink placed at its end. Analytic expressions of energy transfer efficiency are obtained in the high power and low power source limit. Parameters of a photosynthetic system are used as examples to illustrate the results.

\end{abstract}

%\pacs{05.70.Ln,05.60.-k}
%\keywords{Molecular chain;Exciton relaxation;Energy transfer efficiency;Nonequilibrium steady state}

\maketitle

%%%%%%%%%%%%%%%%%%%%%%%%%%
%       Section          %
%%%%%%%%%%%%%%%%%%%%%%%%%%
\section{Introduction}

We study excitation relaxation and excitation energy transfer in molecular chain as an open quantum system \cite{Breuer}.
Excitations can be introduced to a chain by external source through optical absorption. As the excitations transfer through the chain via intersite coupling, they couple to phonons produced by vibrational motion of sites to form collective states in the chain \cite{Kenkre82,May11,Jang20}.
This model can describe the transfer of excitation energy in biological systems, such as in photosynthetic complexes \cite{Leegwater97,Adolphs06,Mohseni08,Olaya08,IshizakiPNAS09,IshizakiJCP09a,[{}][{, and references therein.}]Jang18} and $\al$-helix protein \cite{Davydov79,Davydov90}, in photovoltaic devices \cite{Jang20}, organic semiconductor \cite{Coropceanu07}, and quantum networks \cite{Maier19,Plenio08,Chin10}.

In previous works, because of the structural complications in natural systems, such as photosynthetic systems \cite{Adolphs06}, it was often more convenient to carry out numerical studies \cite{Mohseni08,Olaya08,Plenio08,IshizakiPNAS09,IshizakiJCP09a,Chin10,Plenio12,Jang18}. There were analytic results obtained by considering pure dephasing process from the viewpoint of kinetic networks \cite{Cao09,*Cao20}. It was also found that environmental noise can enhance the transport of energy \cite{Plenio08,Chin10,Plenio12} in these models.

Much effort had also been made to elucidate the role of long-lived quantum coherence in energy transport \cite{Engel07,Panitchayangkoon10,IshizakiPNAS09,Plenio12}, though the interpretations of the results were controversial \cite{Kassal13,[{}][{, and references therein.}]Duan17}. The origin of long-lived coherence was found to depend crucially on the coherent superpositions between exciton and vibrational degrees of freedom \cite{Pachon11,Christensson12}.

In this work, we consider excitation relaxation in amplitude damping or population relaxation process.
Using the usual methods in open quantum systems \cite{Breuer,May11}, we obtain analytic expression of the transition rate between exciton modes in chains with arbitrary number of sites through a few simplifying assumptions, such as neglecting the static disorder in the site energy, and assuming that the sites and phonons are coupled weakly with small inhomogeneity. These assumptions are usually not valid in natural systems such as photosynthetic systems. Static disorder can be neglected, for example, in fabricated systems, where atoms or molecules can be arranged in a more regular pattern, and respond to a more uniform environment.

Equipped with a better understanding of the excitation relaxation process, we investigate energy transfer in the chain at steady state.
In the weak coupling and Markovian limit, coherence components influence the dynamics during the transient only.
We thus focus on the rate equations of exciton occupation numbers. These equations turn out to be similar to those discovered by Fr\"{o}hlich to explain energy storage and transfer in biological systems  \cite{Frohlich68,*Frohlich68b}.

A special feature in the time evolution of exciton occupation number is caused by the existence of nonlinear terms in the rate equations \cite{Frohlich68,Frohlich68b,Mills83}. The nonlinearities affect the final distribution of the occupation numbers significantly when the chain is energized by a sufficiently high power source, i.e., the excitations concentrate to the lowest energy level at steady state. We are able to obtain approximate analytic expressions of the energy transfer efficiency under a high power source based on these results, and under low power source through some general considerations.

There are two ways in which the transfer of excitation energy through a chain could be considered.
Most of the studies used a transient setting \cite{Mohseni08,IshizakiPNAS09,IshizakiJCP09a,Plenio12,Cao09,*Cao20,Plenio08,Cao09,*Cao20,Chin10}, in which one excitation is introduced to the system as an initial condition. This is equivalent to a weak source. The time evolution of the system is then followed until the excitation is finally lost from the chain.

In the second setting which is closer to actual setups, a continuous flow of energy is supplied to the chain and eventually a steady state is achieved. It was shown that the difference in the efficiency between the transient process and the steady state is very small \cite{Jesenko13}. Therefore, it is appropriate to consider energy transfer at steady state.

We also clarify the influence of various parameters on the efficiency of energy transfer through the chain.
In our model which assumes weak coupling between the sites and phonon bath as well as negligible memory effects, internal parameters related to the chain and phonon affect the energy transfer weakly during the transient. Efficiency of energy transfer at steady state is mostly determined by parameters external to the chain and phonon.

Here is an outline of our discussions. We begin in Sec.~\ref{Secsys} with a summary of the procedures required to bring the Hamiltonian of the system into a suitable form in the exciton basis for subsequent analysis. The reduced dynamics is then obtained in Sec.~\ref{secQME}. It has transition rates that can be expressed in closed form owing to a few simplifying assumptions. Their behaviours in chains with large number of sites are discussed. We then obtain the rate equations of exciton occupation numbers in Sec.~\ref{secPop}. External energy source and sink are introduced to enable energy transfer through the chain. The form of the nonequilibrium steady state of the rate equations is also given. In Sec.~\ref{secSrc}, analytic expression of the occupation numbers can be obtained at high power source based on the results of previous works. The distribution of the site occupation numbers at steady state can then be worked out. In Sec.~\ref{secEET}, we obtain the efficiency of energy transfer at steady state through conservation of energy. The special behaviour of the occupation numbers at high power source leads to an analytic expression of efficiency. The efficiency at low power source can also be deduced. We then numerically study the effects of the various parameters on efficiency in Sec.~\ref{secEeta}, and clarify their interconnections based on our understanding of the relaxation dynamics. We conclude our discussions with a short summary of the work. Some of the identities and technical details are presented in the appendices.

%%%%%%%%%%%%%%%%%%%%%%%%%%
%       Section          %
%%%%%%%%%%%%%%%%%%%%%%%%%%
\section{Molecular chain coupled to phonon}
\label{Secsys}

We consider a chain of $\ell$ oscillators located at equal interval from each other. Their coordinates are labeled by $x=1,2,3,\cdots, \ell$.
The Hamiltonian of the system is
%%%
\begin{align}   \label{H}
    H&=H_0+\sum_q \w_q \bdg_q b_q+V\,,\\
    H_0&= \siL \w_0 \adg_x a_x+J\siLm \big(\adg_x a_{x+1}+a_x\adg_{x+1}\big)\,,\label{Hch}
\end{align}
%%%
where we use the units $\hbar=c=1$. $\adg_x$ and $a_x$ denote the creation and annihilation operators of excitation at site-$x$. $b^\dg_q$ and $b_q$ are the corresponding operators of the phonon field. $\w_0$ and $\w_q$ are the frequency or energy of the oscillators and phonon modes, respectively. $J$ is the intersite coupling constant. The number operator of the excitation is coupled to the position operator of phonon linearly,
%%%
\begin{align}   \label{V}
    V&= \siL \sum_q \w_q \chi^{(x)}_q \adg_x a_x \big( b_{q}+\bdg_q\big)\,,
\end{align}
%%%
where $\chi^{(x)}_q$ denotes a dimensionless real coupling strength.

%This is a model of Frenkel excitons \cite{Brown86}, in which excitations are created by optical absorption at the sites and are subsequently transported to neighbouring sites through intersite coupling. Furthermore, the excitations also couple to phonon created by vibrations of each site to form a collective states in the chain. The model describes the transport of energy in photosynthetic systems \cite{Adolphs06,IshizakiJCP09a,Mohseni08}.

By a unitary transformation we can turn the site-phonon interaction into a form involving the difference in the coupling strength between neighbouring sites \cite{May11,Tay14}. This permits us to consider inhomogeneity in the coupling strength. The details were already worked out in Ref.~\cite{Tay14}. Here we will quote the main results that are relevant to our discussions.

We first apply the unitary transformation \cite{Yarkony76,Brown86,May11}
%%%
\begin{align}   \label{U1}
    U=\exp\bigg(-\sum_q \sum^\ell_{x=1}   \chi_q^{(x)} \adg_x a_x (b_{q}-\bdg_q)\bigg)
\end{align}
%%%
on $H$. Terms involving the difference in the coupling between neighbouring sites $\chi^{(x+1)}_q-\chi^{(x)}_q$ will emerge in the resulting expressions. We assume that the difference can be parameterized by a parameter that describes the degree of inhomogeneity $\eta$, also called site-symmetry in Ref.~\cite{Tay14}, in a site-independent form,
%%%
\begin{align}   \label{chij+ve}
    \chi_q^{(x+1)}-\chi_q^{(x)}&= \eta \chi_q
\end{align}
%%%
for all $x$.
By assuming a small and constant inhomogeneity, we expand the resulting expressions in powers of $\eta$, and keep terms linear in $\eta$ to yield the following Hamiltonian \cite{Tay14},
%%%
\begin{align}   \label{H'}
    H'&=H'_0+\sum_q \w_q \bdg_q b_q+V'\,,\\
    H'_0&=\siL  \w'_x \adg_x a_x+J\sum_{x=1}^{\ell-1} (\adg_x a_{x+1}+\adg_{x+1} a_x)\,,\label{H0'}\\
    \w'_x&=\w_0-\sum_q\w_q(\chi^{(x)}_q)^2\,. \label{wx'}
\end{align}
%%%
The correction term to $\w_0$ in \Eq{wx'} is also called reorganization energy \cite{May11}. Under the assumption \eqref{chij+ve}, the interaction becomes
%%%
\begin{align}   \label{V'}
   V'=\eta' \sum_q \siL \w_0 \chi_q &(\adg_x a_{x+1}-\adg_{x+1} a_x)(b_{q}-\bdg_q)\,,
\end{align}
%%%
where $\eta'\equiv \eta J/\w_0$
is a dimensionless parameter. As in most cases $J<\w_0$, later on we will use $\eta'$ as a perturbation expansion parameter when we consider the reduced dynamics of the chain.
We note that we have dropped from $V'$ a quartic term in excitation operators \cite{Tay14}. For small number of excitations, this term is negligible. However, when there is a large number of excitations to the extent that divergence occurs in the cubic term \eqref{V'}, the quartic terms has to be included to the interaction to avoid the divergence, see Ref.~\cite{Nardecchia18} and references therein for details. Another situation in which the quartic terms have to be considered is when the coupling to phonon bath is homogeneous so that $\chi^{(x+1)}_q-\chi^{(x)}_q$ vanishes.

Before we obtain the reduced dynamics of the chain, we first diagonalize $H'_0$ by introducing exciton operators \cite{May11} where $\mu=1,2,\cdots,\ell$,
%%%
\begin{align}   \label{A}
    A_\mu&=\sqrt{\frac{2}{\ell+1}}\siL \sin\left(k_\mu x\right) \, a_x\,,\\
    k_\mu&\equiv \frac{\pi\mu}{\ell+1}\,.\label{k}
\end{align}
%%%
and its hermitian conjugate.
The inverse of \Eq{A} is given in \Eq{ax}.
The exciton operators satisfy the commutation relation $[A_\mu,\Adg_\nu]=\del_\mn$.
To bring $H'_0$ into a diagonalized form, we assume that the correction term to the bare energy $\w_0$ in \Eq{wx'} is negligible due to weak coupling between the sites and phonon. Otherwise, coupling terms involving operators of adjacent sites cannot be diagonalized, see the details in App.~\ref{AppDiscSin}. Adopting this assumption, we obtain the Hamiltonian of the chain in exciton basis
%%%
\begin{align}   \label{H0A}
    H'_0&=\smL \w_\mu \Adg_\mu A_\mu\,,
\end{align}
%%%
with exciton energy
%%%
\begin{align}   \label{wb}
    \w_\mu&\equiv \w_0+2J\cos k_\mu \,.
\end{align}
%%%
Notice that contrary to usual notation, the index $\mu=1,2,\cdots,\ell$ is arranged in a decreasing order of exciton energy, i.e., $\ell$-mode is the lowest energy level of excitons.
%The amplitude of each exciton as a function of $x$ \eqref{A} is reminiscent to the stationary waves of a string fixed at both ends, $x=0,  \ell+1$, with wave number $k_\mu$. When the energy of the excitons reduces from $\w_1$ to $\w_\ell$, the square of the amplitude approaches a more uniform distribution.

In the exciton basis the interaction becomes
%%%
\begin{align}   \label{V'exc}
    V'&=\eta'  \sum_q  \w_0\chi_q \smn c_\mn (L^\dg_\mn-L_\mn)( b_{-q} -\bdg_q) \,,
\end{align}
%%%
where exciton indices are arranged according to the order $\mu<\nu$. The summation symbol with a prime abbreviates a double summation over $\mu$ and $\nu$ excluding $\mu= \nu$ terms,
%%%
\begin{align}   \label{sum}
    \smn&\equiv \sum_{\mu=1}^{\ell-1}\sum_{\nu=\mu+1}^\ell \,.
\end{align}
%%%
The exciton raising operator
%%%
\begin{align}   \label{Lmn}
    L_\mn ^\dg&\equiv \Adg_\mu A_\nu
\end{align}
%%%
creates an exciton of energy $\w_\mu$ while simultaneously annihilates another one of energy $\w_\nu$. Its hermitian conjugate is the lowering operator $L_\mn =\Adg_\nu A_\mu$.

The exciton basis introduces a new coefficient $c_\mn$ to $V'$,
%%%
\begin{align}   \label{cpm}
    c_{\mu\nu}&\equiv \frac{2}{\ell+1}\sum_{x=1}^{\ell-1}
    \big[\sin(k_\mu x)\sin(k_\nu (x+1))\nm
    &\qquad\qquad\qquad\quad-\sin(k_\mu (x+1))\sin(k_\nu x)\big]\,.
\end{align}
%%%
The sum over site index can be carried out exactly. We first simplify the expression by combining the products of sine functions. Then we sum over site-$x$ using identities \eqref{Propcos} and \eqref{Propsin} to obtain
%%%
\begin{align}   \label{cmn+}
    c_{\mu\nu}&=\left\{\begin{array}{ccl}
                  \ds \frac{4}{\ell+1} \frac{\sin k_\nu\sin k_\mu}{\cos k_\nu-\cos k_\mu}\,,  & &\nu-\mu=\text{odd only}\,, \\\\
                  0\,, &  & \text{otherwise}\,,
                \end{array}\right.
\end{align}
%%%
where $\mu<\nu$.
It shows that excitons with odd indices are coupled only to excitons with even indices, and vice versa.
There are $\ell^2/4$ or $(\ell^2-1)/4$ pairs of coupled excitons for even $\ell$ or odd $\ell$, respectively.
The energy gap between two exciton levels is
%%%
\begin{align}   \label{wbmn}
        \w_\mn&\equiv \w_\mu-\w_\nu = 2J(\cos k_\mu-\cos k_\nu)\,.
\end{align}
%%%
Hence, $c_\mn$ is inversely proportional to the energy gap.
Pair of excitons with adjacent indices are coupled most strongly to phonon.
In particular, maximum coupling occurs between pair lying around the center of the exciton spectrum $\ell/2$, when the numerator of $c_\mn$ is also largest.
In the limit of very long chain $\ell\gg1$, $c_\mn\rightarrow -4/\pi$ approaches its maximum magnitude.

%To obtain the reduced dynamics of the system, we need to introduce specific models on the difference in the site-coupling to the phonon field $\chi_q^{(x)}-\chi_q^{(x+1)}$ between neighbouring sites. We postpone the details of the models to Sec.~\ref{secMod}. Here we just assume that the difference in the site-coupling to the phonon field is proportional to a small parameter $\eta$, cf.~\Eq{chij+ve}, with respect to which perturbation is carried out. Then, the group of trigonometric functions can be condensed into a single model-dependent parameter $c_{\mn}$ \eqref{cpm} to yield a simpler expression of
%%%
%\begin{align}   \label{VLc}
%    V'&=\eta J\sum_q  \w_q \chi_q \smn c_{\mu\nu}
%    (L^\dg_{\mu\nu}-L_{\mu\nu})( b_q -\bdg_q) \,,
%\end{align}
%%%

%%%%%%%%%%%%%%%%%%%%%%%%%%
%       Section         %
%%%%%%%%%%%%%%%%%%%%%%%%%%i
\section{Reduced dynamics of molecular chain}
\label{secQME}

Assuming a small inhomogeneity $\eta$ and a weak coupling between the sites and phonon, we apply the rotating-wave approximation and the Markovian approximation \cite{Breuer} to obtain the quantum master equation for the reduced density operator of chain $\rho$, using $\eta'$ in $V'$ \eqref{V'} as a perturbation parameter. The chain is in contact with a phonon bath in thermal equilibrium obeying the Bose-Einstein distribution
%%%
\begin{align} \label{nbmn}
        n^\th_\mn&\equiv \frac{1}{e^{\bt \w_\mn}-1}\,,
\end{align}
%%%
with the inverse temperature $\bt\equiv 1/(k_B T)$.

The time evolution equation of $\rho$ is
%%%
\begin{align} \label{drhodt}
    \frac{\d \rho}{\d t}\bigg|_\ch&=-K_\ch\rho\,,
\end{align}
%%%
where $K_\ch\equiv K_0+K_\text{d}$, in which
%%%
\begin{align}   \label{K0}
        K_0\rho&\equiv i[H'_0,\rho]\,,\\
        K_\text{d}\rho&\equiv-\frac{1}{2}\smn
                \big(\Gam_\mn n^\th_\mn R_\mn \rho +\Gam_\mn (n^\th_\mn+1) S_\mn \rho\big)\,.\label{Kd}
\end{align}
%%%
%The equation conserves probability $\d\tr(\rho)/\d t=0$.
The operator $R_\mn$ and $S_\mn$ have the Kossakowski-Lindblad form
%%%
\begin{align}
        R_\mn\rho&\equiv 2L_\mn ^\dg\rho L_\mn-L_\mn \Ldg_\mn\rho-\rho L_\mn \Ldg_\mn\,,\\
        S_\mn\rho&\equiv 2L_\mn\rho L_\mn ^\dg -\Ldg_\mn L_\mn \rho-\rho \Ldg_\mn L_\mn\,.
\end{align}
%%%
We have omitted a renormalization to the frequency of the exciton modes \cite{Tay13,Tay14}.
The transition rate between two exciton levels can be written in the form
%%%
\begin{align}   \label{Gam}
        \Gam_\mn&\equiv c_\mn^2 \gam^{(2)}_\mn\,.
\end{align}
%%%
In Ref.~\cite{Tay14}, it was shown that the relaxation rate of a dimer in a phonon bath
%%%
\begin{align}   \label{gam2}
        \gam^{(2)}_\mn&\equiv\left(\eta\frac{J}{\w_\mn}\right)^2 \gam_\text{d}
\end{align}
%%%
is slowed down by the factor in the round bracket over the dephasing rate of a single site in contact with a phonon bath
%%%
\begin{align}   \label{gamd}
        \gam_\text{d}&\equiv 2\pi\sum_q \w_q^2 (\chi_q)^2 \del(\w_q-\w_\mn)\,.
\end{align}
%%%
In the following we assume that $\gam_\text{d}$ is constant to simplify our analysis.
The slow down results in a longer coherence lifetime.
When chains longer than two sites are considered, the slow down in the transition rate is still true for shorter chains. However, as the number of sites increases, the transition rate starts to increase and eventually exceeds $\gam_\text{d}$ to result in shorter coherence lifetime.

We can understand the decrease in the coherence lifetime better by analyzing the behaviour of transition rate as a function of the number of sites $\ell$.
The transition rate has a complicated dependence on the pair of coupled exciton indices
%%%
\begin{align}   \label{gamc+}
        \Gam_\mn&=\frac{4\eta^2 \gam_d}{(\ell+1)^2}\frac{\sin^2k_\mu \sin^2k_\nu}{(\cos k_\mu-\cos k_\nu)^4}\,,
        &\nu-\mu=\text{ odd only}\,,
\end{align}
%%%
and 0 otherwise. It is inversely proportional to the fourth power of the energy gap between exciton levels \eqref{wbmn}.

The transition rates can be separated into series each containing rates with similar order of magnitude. The series is labelled by an odd integer, $m=\nu-\mu=1,3,5,\cdots$. As indicated at the end of Sec.~\ref{Secsys}, the transition rate is highest in the $m=1$ series when the energy gap $\w_\mn$ is smallest and the magnitude of $c_\mn$ largest.
The rate decreases rapidly as $m$ increases. We demonstrate this fact by an estimate of the ratio between $\Gam_\mn$ of the 1-series to the next few series as follows.
The maximum of each series occurs between the pair of indices $\mu=\ell/2$ and $\nu=\ell/2+m$. After substituting them into \Eq{gamc+}, we expand the expression in powers of $m/\ell$ and consider the long chain limit $\ell\gg1$. The leading term in the expansion is
%%%
\begin{align} \label{GamlargeL}
        \Gam_{\ell/2,\ell/2+m}&\approx \frac{4\eta^2\gam_d}{\pi^4}\cdot\frac{\ell^2}{m^4}\,, \qquad m/\ell \ll 1\,.
\end{align}
%%%
Consequently, the ratio of transition rates between the 1- and the $m$-series is
%%%
\begin{align} \label{Gamratio}
        \frac{\Gam_{\ell/2,\ell/2+m}}{\Gam_{\ell/2,\ell/2+1}}
        &\approx \frac{1}{m^4}\,, \qquad m/\ell\ll 1\,,
\end{align}
%%%
which decreases rapidly with an increase in $m$.

This implies that the exciton relaxation dynamics in long chains is dominated by the 1-series, which involves transitions between excitons nearest in energy level. At low temperature $n^\th_\mn\approx0$, excitons of higher energy level cascade down to the lowest level $\nu=\ell$, which acts like a metastable state before the excitation is lost to a sink or through radiation.

%We note that a different type of superradiance transition occurs in one-dimensional nanostructures \cite{Celardo09}. There, a few states become short-lived, while other long-lived states become effectively decoupled from the relaxation dynamics. In contrast, in the reduced dynamics considered here, at low temperature all higher level excitons decay rapidly to a single state at the lowest energy.

In shorter chain, the transition rates remain small $\Gam_\mn/\gam_\text{d}<1$. This means that the coherence lifetime between exciton levels will survive longer as in dimer \cite{Tay14}. For example, when $\eta=0.1$, $\Gam_\mn/\gam_\text{d}<1$ for up to $\ell=48$.
When the number of sites continue to increase, the ratio $\Gam_\mn/\gam_\text{d}$ eventually exceed 1, resulting in rapid relaxation compared to dephasing in single site.

%%%%%%%%%%%%%%%%%%%%%%%%%%
%       Section         %
%%%%%%%%%%%%%%%%%%%%%%%%%%i
\section{Rate equation of exciton occupation number}
\label{secPop}

We next consider the average number of excitations and the coherence between them.
From now on, we arrange three indices, $\mu,\sig$ and $\nu$, in the order $\mu<\sig<\nu$.
Denoting the trace of an operator over the reduced density operator by $\<O\>\equiv\tr\big(O\rho\big)$, the occupation number and the correlation function of exciton operators are
%%%
\begin{align}   \label{Nmu}
    n_\sig&\equiv \<\Adg_\sig A_\sig\>\,,\\
    n_\ms&\equiv \<\Adg_\mu A_\sig\>\,,\\
    m_\ms&\equiv \<A_\mu A_\sig\>\,, \label{mms}
\end{align}
%%%
together with their complex conjugates.
%In the following discussion, we will assume that correlation functions of the type $\<A_\mu A_\sig\>$ and its complex conjugate is zero.
To obtain the time evolution of \Eqs{Nmu}-\eqref{mms}, we need to trace operators quadratic in $A$s over \Eq{drhodt}. The expression can be reduced to a form involving correlation function of quartic operator. We approximate them by the products of correlation function of quadratic operator, for example,
%%%
\begin{align}   \label{aveAA}
\<\Adg_\mu \Adg_\nu A_\sig A_\kap\>&=m^*_\mn m_\sk+n_\ms n^*_\kn+n_\mk n^*_\sn\,,
%\<\Adg_\mu A_\sig\>\<\Adg_\nu A_\tau\>+\<\Adg_\mu A_\tau\>\<\Adg_\nu A_\sig\>\nm
%&\quad+\<\Adg_\mu \Adg_\nu\>\<A_\sig A_\tau\> \,,
\end{align}
%%%
where we assume that the trace of odd number products of $A$ and $\Adg$ have zero trace.
The complete set of rate equations are given in App.~\ref{AppCompRedDyn}.

The correlation functions satisfy the following Schwartz inequality \cite{Englert03,*Tay19b}
%%%
\begin{align}   \label{Schwartzn}
    |n_\mn|^2&\leq n_\mu n_\nu  \,,\\
    |m_\mn|^2&\leq n_\mu(n_\nu+1) \,. \label{Schwartzm}
\end{align}
%%%
Consequently, their magnitudes are constraint by their diagonal counterparts.
Numerical studies on the complete set of rate equations also suggest that the correlation components indeed affect the evolution of the occupation numbers only weakly.
Moreover, they vanish eventually in the long time limit, thus playing no role in the steady state.

For these reasons, we will drop the correlation terms from the rate equation and consider the time evolution in occupation numbers only,
%%%
\begin{align}   \label{ndt}
    \frac{dn_\sig}{dt}\bigg|_\ch&=
    \sum_{\mu=1}^{\sig-1}\Gam_\ms\big[n^\th_\ms(n_\mu-n_\sig)
            + n_\mu(1+n_\sig)\big] \nm
    &+\sum_{\nu=\sig+1}^\ell\Gam_\sn\big[n^\th_\sn(n_\nu-n_\sig)
    -n_\sig(1+ n_\nu)\big] \,.
\end{align}
%%%
%where we use the notations
%%%
%\begin{align}
%        \sum_{\mu<\sig}&\equiv %\sum_{\mu=1}^{\sig-1}\,, &\quad %\sum_{\nu>\sig}&\equiv \sum_{\nu=\sig+1}^\ell\,,
%\end{align}
%%%
Notice that terms containing the influence of phonon bath, such as $n^\th_{\ms}$ and $n^\th_{\sn}$, induce transitions of excitons between a pair of levels in both directions $\mu\leftrightarrow\sig\leftrightarrow\nu$. The rest are the ``spontaneous" emission terms, which permit transitions directed towards lower energy levels only $\mu\rightarrow\sig\rightarrow\nu$. The spontaneous terms contain nonlinear products of exciton occupation numbers, for instance, $n_\mu n_\sig$ and $n_\sig n_\nu$. When the power of an energy source supplied to the chain is beyond a certain value, these nonlinear terms induce the majority of excitations to stay in the lowest energy mode at steady state. This phenomenon is called Bose-Einstein condensation in biological systems \cite{Frohlich68,Frohlich68b,Nardecchia18}. We will discuss the distribution of occupation number at steady state in Sec.~\ref{secSrc} after we introduce external sink and source to the chain.

%With the addition of source and bath, it turns out that \Eq{ndt} leads to long range coherence in biological systems \cite{Frohlich68,Wu77,Mills83,Tuszyinski84}. When the energy supplied by the source is beyond certain rate, a steady state will be reached in which most of the excitations is concentrated in the lowest energy mode. The phenomenon also manifests as collective oscillations leading to classical phonon condensation in protein \cite{Nardecchia18}. \Eq{ndt} differs from the usually considered evolution with the addition of the last nonlinear terms of correlation functions in the second and third line.

The total number of exciton occupation numbers in the chain
%%%
\begin{align}   \label{N}
    N\equiv \ssL n_\sig\,,
\end{align}
%%%
is a constant of motion, by virtue of
%%%
\begin{align}   \label{ntch0}
       \ssL\frac{dn_\sig}{dt}\bigg|_\ch=0\,.
\end{align}
%%%
%In the limit $t\rightarrow\infty$, the cross correlation functions as well as their complex conjugates vanish, $n_\mn, m_\mn\rightarrow 0$, whereas
The stationary state of each mode is
%%%
\begin{align}   \label{NbSS}
    \nb_\sig\big|_\ch&=\frac{1}{\exp\big[\bt(\w_\sig-\mu^\text{c})\big]-1}\,,
\end{align}
%%%
with a constant chemical potential $\mu^\text{c}$. Its value can be determined through \Eq{N}. From now on, we use ``bar" to denote quantities at steady state.

Later, it will be interesting to consider the exciton occupation number in the site basis, given by
%%%
\begin{align}   \label{nsite}
    n^{(\st)}_x&=\tr\big(\adg_x a_x \rho\big)
%        &=\frac{2}{\ell+1}\smL i \sum_{\nu=1}^{\ell}
%        \sin(k_\mu x)\sin(k_\nu x) n_\mn\,.
    =\frac{2}{\ell+1}\smL
        \sin^2(k_\mu x) n_\mu\,,
\end{align}
%%%
where we have dropped the correlation component $n_\mn$ which vanishes in the steady state.
Notice that the site occupation number is symmetric with respect to the center of the chain $(\ell+1)/2$, for \Eq{nsite} gives $n^{(\st)}_x=n^{(\st)}_{\ell+1-x}$, see \Fig{fig1} for examples of excitation profile along the chain.
By means of the identity \eqref{Snn}, we verify that the total site occupation number is the same constant of motion as in \Eq{N},
%%%
\begin{align}   \label{sumnsite}
    \siL n^{(\st)}_x&=N\,.
\end{align}
%%%

Let us now couple external sink and energy source to the chain.
In App.~\ref{AppExt}, we discuss how this could be done.
When a field is coupled to the chain through an interaction linear in both the site and field operator \eqref{sourceH}, the resulting rate equation \eqref{sknT} has a component that functions like a source, whereas the other component acts like a sink.
To better separate the contribution of the two components, we couple ``pure" source and ``pure" sink to the chain, see App.~\ref{AppExt} for the details.

A ``pure" sink coupled to site-$z$ of the chain will give the following contribution to the rate equation
%%%i
\begin{align}   \label{SSK}
        \frac{dn_\sig}{dt}\bigg|_\si&=-\al^{(z)}_\sig \gam_\s n_\sig\,,
\end{align}
%%%
where $\gam_\s$ is the trapping rate of the sink. The trapping power is distributed over all modes according to the weight
%%%
\begin{align}   \label{alz}
    \al^{(z)}_\sig\equiv\frac{2}{\ell+1} \sin^2(k_\sig z)\,,
\end{align}
%%%
which satisfy $\sum_{\sig=1}^\ell \al^{(z)}_\sig=1$.
It shows that with a sink coupled to the end of the chain at $z=\ell$, exciton with energy closer to the center of the energy spectrum will experience the fastest trapping rate.

A ``pure" source that introduces $s$ excitations per unit time to the chain through site-1 can be described by adding the following term to the rate equation
%%%
\begin{align}   \label{sourcent}
        \frac{dn_\sig}{dt}\bigg|_\src
        &=s_\sig\,,\\
        s_\sig&\equiv \al^{(1)}_\sig s\,, \label{ssig}
\end{align}
%%%
see \Eq{extSrc}. As defined in \Eq{alz}, $\al^{(1)}_\sig$ is the fraction of excitations channeled to the $\sig$-mode exciton.
The source could be a radiation field that excites the chain, such as in photosynthetic systems. Creating an excitation at site-1 from its ground state requires an energy of $\w_0$. Hence, the power of the source is $\w_0s$.
%On the other hand, a uniform source that evenly distributes its power to all the modes has a partial power of $s_\sig=s/\ell$.

Radiation emitted following the relaxation of an excitation per unit of time to the ground state leads to loss of energy. In a similar way to the sink, radiation loss can be described by adding a term
%%%
\begin{align}   \label{dnrad}
        \frac{dn_\sig}{dt}\bigg|_\text{rad}&=-\gam_\rad n_\sig
\end{align}
%%%
to the rate equation. Here, we assume that all sites equally radiate, thus a constant radiation rate $\gam_\rad$ for all modes.

Combining the various contributions to the energy exchange process in the chain, the final rate equation we consider is
%%%
\begin{align}   \label{dndttot}
        \frac{dn_\sig}{dt}\bigg|_\tot&=
        \frac{dn_\sig}{dt}\bigg|_\src
        +\frac{dn_\sig}{dt}\bigg|_\si
        +\frac{dn_\sig}{dt}\bigg|_\text{rad}
        +\frac{dn_\sig}{dt}\bigg|_\ch \nm
        &=s_\sig- \eps^{(z)}_\sig n_\sig
        +\frac{dn_\sig}{dt}\bigg|_\ch\,,\\
        \eps^{(z)}_\sig&\equiv \al_\sig^{(z)} \gam_\s+\gam_\rad\,.
\end{align}
%%%
This equation has two clearly separated time scales. We already discussed in Sec.~\ref{secQME} that longer chains have fast transition rates dominated by $\Gam_{\ell/2,\ell/2+1}$ from the 1-series. It provides an estimate of the shorter time scale $\tau_1\sim1/\Gam_{\ell/2,\ell/2+1}$ in the reduced dynamics. The longer time scale $\tau_2$ is provided by the loss through external sink and radiation, with a smaller relaxation rate $\eps^{(z)}_\sig$ \eqref{dndttot}. Hence, the estimate $\tau_2\sim 1/\eps^{(z)}_\sig$.

It turns out that \Eq{dndttot} together with \Eq{ndt} has a similar form to the rate equation of Fr\"ohlich model for biological systems \cite{Frohlich68,Frohlich68b,Wu77,Mills83}.
It should therefore exhibit a phenomenon similar to Bose-Einstein condensation \cite{Frohlich68,Frohlich68b}, where most of the excitations concentrate to the lowest energy level. This occurs when the power of the source is sufficiently high.

Summing the rate equations over all exciton modes at steady state produces an equation that relates the power of the source to the parameters of loss mechanism,
%%%
\begin{align}   \label{sumnt}
    s&=\ssL\eps^{(z)}_\sig \nb_\sig\,,
\end{align}
%%%
by means of \Eq{ntch0}.

The exciton occupation numbers at the steady state have a similar form as \eqref{NbSS},
%%%
\begin{align}   \label{NbSSsrc}
    \nb_\sig&=\frac{1}{\exp\big[\bt(\w_\sig-\mu^\text{c}_\sig)\big]-1}\,,
\end{align}
%%%
except that now different modes have different chemical potentials to account for nonequilibrium steady state \cite{Kondepudi14}.
The solutions to the exciton occupation numbers at steady state can be obtained numerically by finding the roots of the coupled nonlinear equations \eqref{dndttot} together with \Eq{ndt}.
They also satisfy the consistency condition \eqref{sumnt}.

%%%%%%%%%%%%%%%%%%%%%%%i%%%
%        Section         %
%%%%%%%%%%%%%%%%%%%%%%%%%%
\section{Occupation numbers at steady state under high power source}
\label{secSrc}

A special feature of the reduced dynamics is the existence of nonlinear terms in the rate equation. The origin of these terms can be traced back to the cubic coupling between the site number operator $\adg_x a_x$ and the phonon field. Similar rate equation was introduced to explain the storage and transfer of energy in biological systems \cite{Frohlich68,Frohlich68b}. It was predicted that when the source power exceeds certain value, most of the excitations condenses to the lowest energy level, giving rise to a coherent oscillations of the entire chain. Without the nonlinear terms, the occupation numbers will distribute more uniformly across all the modes according to temperature.
This condensation was recently reported in protein as a classical phenomenon \cite{Nardecchia18}.

In the special case of sufficiently high power source, the steady state exciton occupation numbers can be approximated analytically \cite{Frohlich68,Frohlich68b,Mills83}. As most of the excitations condenses to the lowest level $\ell$-mode at high power source, the chemical potential for the lowest mode $\mu^\text{c}_\ell$ has to approach $\w_\ell$ in order to support a large excitation in this mode \eqref{NbSSsrc}. It happens that the chemical potentials of other modes $\mu^\text{c}_\sig$ also approach $\w_\ell$ \cite{Mills83}. Therefore, we can approximate them by
%%%
\begin{align}   \label{mucn}
        \mu^\text{c}_\ell&\approx\w_\ell(1-\del_\ell)\,,\\
        \mu^\text{c}_\sig&\approx\w_\ell(1+\del_\sig)\,,\qquad \sig<\ell\,,
\end{align}
%%%
where the $\del$s are small quantities. We estimate them in the following paragraphs.

When we regard the occupation number as a function of source power $s$, in the high $s$ limit the occupation number in the lowest mode is linear in $s$, whereas the occupation numbers in other modes start at $O(s^0)$ \cite{Frohlich68,Frohlich68b,Mills83}. By expanding the occupation number in powers of $s$, then substituting them into \Eq{dndttot} together with \Eq{ndt} and extracting terms of the same order in $s$, we obtain the coefficients of expansion \cite{Mills83}. As a result,
%%%
\begin{align}   \label{nLs}
    \nb_\ell&= s/\eps^{(z)}_\ell+O(s^0)\,,\\
    \nb_\sig&=\left(1+\frac{\al^{(1)}_\sig\eps^{(z)}_\ell}{\Gam_{\sig\ell}}\right)
    \frac{1}{e^{\bt\w_{\sig\ell}}-1}+O(s^{-1})\,,\qquad \sig<\ell\,. \label{nsigs}
\end{align}
%%%

On the other hand, by expanding the occupation number \eqref{NbSSsrc} in powers of $\del$s, we can write them in a similar form to \Eqs{nLs} and \eqref{nsigs}. Upon comparing both sets of expressions, we deduce that
%%%
\begin{align}   \label{delell}
        \del_\ell&\approx\frac{kT}{\w_\ell}\cdot\frac{\eps^{(z)}_\ell}{s}\,,\\
        \del_\sig&\approx\frac{kT}{\w_\ell}\cdot
        \frac{\al^{(1)}_\sig \eps^{(z)}_\ell}{\Gam_{\sig\ell}}\,,\qquad \sig<\ell\,.    \label{delsig}
\end{align}
%%%
We should note that we do not apply the high temperature limit in our consideration.

The requirement $\del_\ell\ll 1$ then provides a condition whereby the approximation in \Eqs{nLs} and \eqref{nsigs} should hold during condensation,
%%%
\begin{align}   \label{scond}
       s\gg \frac{kT}{\w_\ell} \eps^{(z)}_\ell \,.
\end{align}
%%%
Imposing the requirement $\del_\sig\ll 1$, we obtain a condition satisfied by the other modes
%%%
\begin{align}   \label{scondsig}
       \Gam_{\sig\ell}\gg \frac{kT}{\w_\ell} \al^{(1)}_\sig \eps^{(z)}_\ell \,,\qquad \sig<\ell\,.
\end{align}
%%%
The approximation breaks down when $\Gam_{\sig\ell}$ \eqref{gamc+} vanishes in certain modes.
When this occurs, we can estimate the occupation number for this mode using the rate equation at the steady state \eqref{dndttot}. To this end, we neglect the $\mu<\sig$ terms which are increasingly smaller, then we solve for $\nb_\sig$ to obtain
%%%
\begin{align}   \label{nsigsSS}
    \nb_\sig&\approx\frac{\ds s_\sig+\sum_{\nu=\sig+1}^{\ell}
    \Gam_{\sig\nu}n^\th_{\sig\nu} \nb_\nu}
    {\ds \eps^{(z)}_\sig+\sum_{\nu=\sig+1}^{\ell}
    \Gam_{\sig\nu}(n^\th_{\sig\nu} +\nb_\nu+1)}\,.
\end{align}
%%%

%%%
\begin{figure}[t]
\centering
\includegraphics[width=3.2in, trim = 5.5cm 11cm 3cm 10cm]{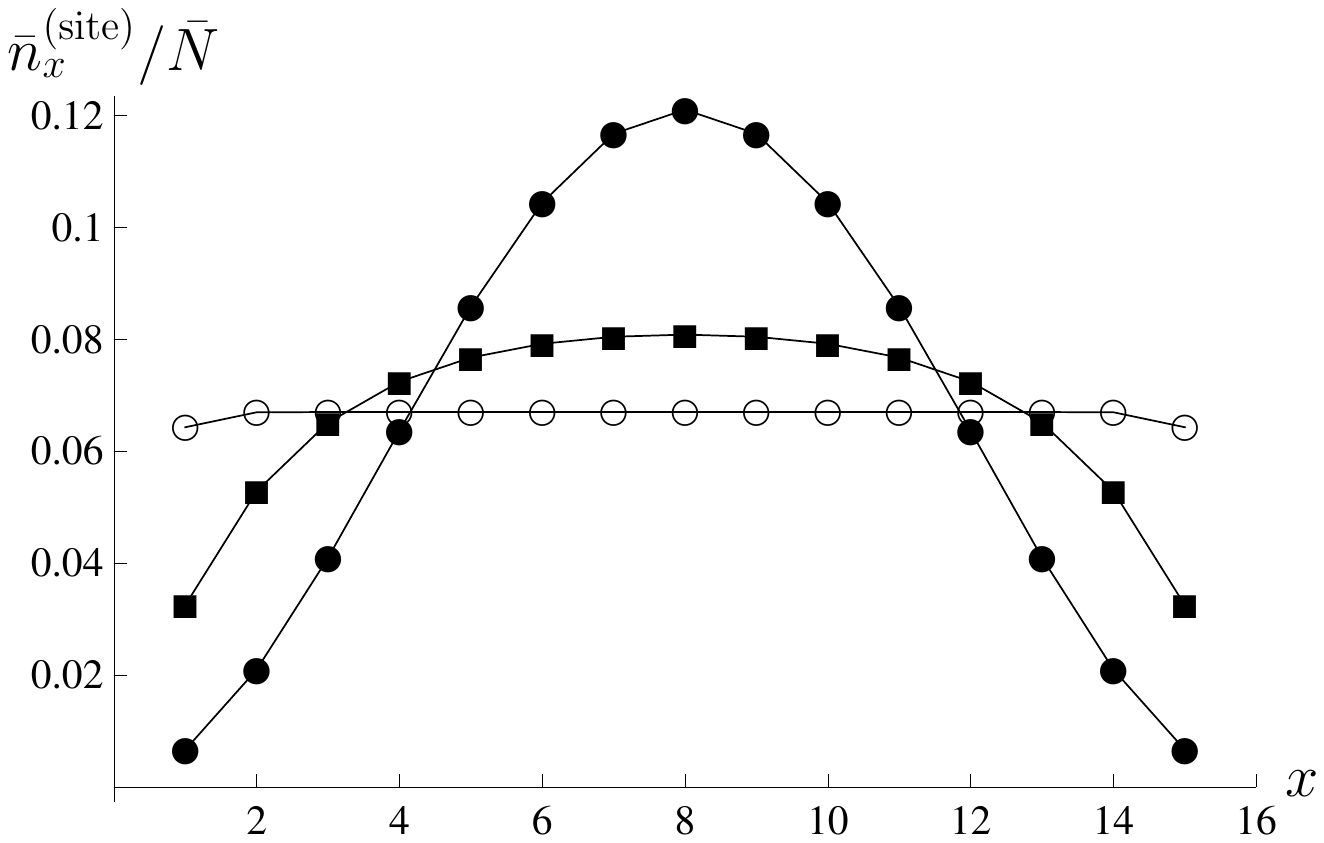}
\caption{$\bar{N}$ is the total number of excitations \eqref{sumnsite} at steady state. We use the set of reference parameters listed at the beginning of Sec.~\ref{secEeta}, where $J=100\,\cmi$, $\w_0=12,500 \,\cmi$, $\gam_\text{d}=20 \text{ ps}^{-1}$, $\gam_\s=1\,\psi$, $\gam_\rad=0.001\text{ ps}^{-1}$and $\eta=0.1$. The symbols denote $\circ=(77,0.001)$, $\bullet=(77,10)$ and $\blacksquare=(300,10)$, where the numbers in brackets denote $(T\text{ K},s\,\psi)$. The sink is prepared at the end of the chain.
}
\label{fig1}
\end{figure}
%%%i

It is interesting to investigate the profile of the normalized site occupation numbers along the chain at steady state.
We make use of the typical values of parameters in photosynthetic system \cite{Adolphs06,Mohseni08} listed in the caption of \Fig{fig1}, see also discussion at the beginning of Sec.~\ref{secEeta}.
For a low power source of $s=0.001\,\psi$ at a temperature of 77 K, the exciton modes are almost evenly excited. The distribution of the occupation number in the site basis is uniform across the chain, depicted by the $\circ$-curve in \Fig{fig1}, where we consider a chain with 15 sites.
Other parameters of the curves are listed in the caption.
With an increase in the power of the source to $s=10\,\psi$ at the same temperature, the excitations condense to the lowest energy mode.
The distribution of the site occupation numbers then approaches the profile of the lowest exciton mode (in the site basis) as depicted by the $\bullet$-curve in \Fig{fig1}. It reminisces the profile of the lowest stationary mode of a vibrating string fixed at both ends.

With the source power fixed at this higher rate, a further increase in the temperature of the phonon bath to 300 K removes excitations trapped in the lowest mode and excites them to higher modes, thus distributing the excitations more evenly among all the modes. We then regain a more uniform distribution ($\blacksquare$-curve) that approaches the $\circ$-curve at even higher temperature.

We can also learn from \Fig{fig1} how high $s$ should be to initiate condensation. In the $\bullet$-curve, about 95\% of the excitations are in the lowest mode, followed by 2\%, 1\% and etc., in subsequent higher modes. With a value of $\del_\ell\approx2.5\times10^{-6}$ and $s=10\,\psi$, it implies that $s$ has to be about $4\times10^5$ times greater than $\eps^{(z)}_\ell kT/\w_\ell$ to fully achieve condensation to the lowest mode. On the other hand, in the $\blacksquare$-curve, condensation is only partially realized. The fraction of excitations in the lowest mode is about 42\%, followed by 20\%, 11\% and etc., in subsequent higher modes. It has $\del_\ell\approx9.7\times10^{-6}$, which means $s$ is about $10^5$ times greater than $\eps^{(z)}_\ell kT/\w_\ell$.

%%%%%%%%%%%%%%%%%%%%%%%%%%
%       Section         %
%%%%%%%%%%%%%%%%%%%%%%%%%%
\section{Energy transfer efficiency at steady state}
\label{secEET}

As already mentioned in Sec.~\ref{secPop}, a source that supplies a constant rate of excitations $s$ to the chain at site-1, channels a rate of energy $\w_0s$ to the chain. This fact is consistent with the sum of the rate of exciton energy over all the modes,
%%%
\begin{align}
       e_\src&\equiv \ssL\w_\sig \frac{dn_\sig}{dt}\bigg|_\src=\ssL\w_\sig s_\sig =\w_0 s\,,
\end{align}
%%%
where the last equality is obtained by simplifying the products of trigonometric functions in $\w_\sig$ \eqref{wb} and $s_\sig$ \eqref{ssig}, followed by using identity \eqref{Propcos}.

When an excitation relaxes to the ground state, its energy can be either collected by the sink as useful energy, or lost to radiation wasted. At steady state, the rate of energy captured by the sink is
%%%
\begin{align}
       e_\si&\equiv -\ssL \w_\sig \frac{d\nb_\sig}{dt}\bigg|_\si
        =\ssL \w_\sig \al^{(z)}_\sig \gam_\s\nb_\sig \,.
\end{align}
%%%
The radiation loss has a similar expression
%%%
\begin{align}
       e_\text{rad}&\equiv -\ssL \w_\sig \frac{d\nb_\sig}{dt}\bigg|_\text{rad}
        =\ssL \w_\sig \gam_\rad \nb_\sig  \,.
\end{align}
%%%
The loss of energy to phonon bath is given by applying energy conservation through the steady state condition, $d\nb_\sig/dt|_\tot=0$, to \Eq{dndttot} to yield
%%%
\begin{align}
       e_\ph&=e_\src-e_\si-e_\text{rad}\,.
\end{align}
%%%
It can be further simplified into a compact form
%%%
\begin{align}
       e_\ph&\equiv -\ssL \w_\sig \frac{d\nb_\sig}{dt}\bigg|_\ch\nm
        &=\smn \w_\mn \Gam_\mn \big[n^\th_\mn(\nb_\mu-\nb_\nu) +\nb_\mu (\nb_\nu+1)\big]\,.
\end{align}
%%%
The sum of $e_\ph$ and $e_\text{rad}$ then amounts to the total rate of energy dissipated by the chain,
%%%
\begin{align}
       e_\diss&\equiv e_\ph+e_\text{rad}
        =e_\src-e_\si\,.
\end{align}
%%%
%The partial energy transfer efficiency through each mode is
%%%
%\begin{align}
%       \eta_e(\sig)&\equiv \frac{e_\si(\sig)}{e_\src(\sig)}\,,
%\end{align}
%%%
Finally, the energy transfer efficiency at steady state is
%%%
\begin{align}
       \eta_e&\equiv \frac{e_\si}{\ds e_\src}=1-\frac{e_\diss}{e_\src}\,,
\end{align}
%%%
which also equals
%%%
\begin{align}   \label{etaEsumsig}
       \eta_e&=\frac{\ds\ssL\frac{\w_\sig}{\w_0}\al^{(z)}_\sig\gam_\s \nb_\sig}{\ds\ssL\eps^{(z)}_\sig \nb_\sig}\,,
\end{align}
%%%
where we use \Eq{sumnt} in place of $s$ in the denominator. This expression has a similar form to the quantum trapping yield obtained in Ref.~\cite{Cao09,*Cao20}. In App.~\ref{Apptranseta} we show that if we consider the efficiency by following the evolution of the system, it approaches \Eq{etaEsumsig} in the long time limit when the steady state is reached.
It was previously shown in Ref.~\cite{Jesenko13} that an initial excitation that drives the reduced dynamics produces almost identical efficiency with continuous excitations provided by a source.

%%%%%%%%%%%%%%%%%%%%%%%%%%
%       Section         %
%%%%%%%%%%%%%%%%%%%%%%%%%%
\subsection{High power source}
\label{secHighs}

When the high power source condition \eqref{scond} is fulfilled, we can approximate $\eta_e$ by keeping contribution from the lowest dominant mode only \eqref{nLs}, to yield
%%%
\begin{align}   \label{etaElarges}
       \eta_e&\approx \frac{\w_\ell}{\w_0} \frac{1}{1+\ds\frac{\gam_\rad}{ \al^{(z)}_\ell\gam_\s}}\,.
\end{align}
%%%
Let us now consider two situations, whether a sink is prepared at the end or at the center of the chain.

(1) Sink is at the end $z=\ell$. In this configuration, we approximate the weight $\al_\ell^{(z)}$ \eqref{alz} by
%%%
\begin{align}   \label{alzl}
       \al_\ell^{(\ell)}&=\frac{2}{\ell+1}\sin^2\left(\frac{\pi}{\ell+1}\right)
        \approx \frac{2\pi^2}{(\ell+1)^3}\,.
\end{align}
%%%
Then, the efficiency has the following expressions in two opposite limits,
%%%
\begin{align}   \label{etazl}
       \eta_e&\approx \frac{\w_\ell}{\w_0}
        \times \left\{\begin{array}{ccc}
        \ds
            1-\frac{(\ell+1)^3}{2\pi^2}\frac{\gam_\rad}{\gam_\s}\,,
            & &\ds \frac{(\ell+1)^3}{2\pi^2}\gam_\rad\ll \gam_\s\,,\\\\
            \ds\frac{2\pi^2\gam_\s}{(\ell+1)^3\gam_\rad}\,,
            & &\ds \frac{(\ell+1)^3}{2\pi^2} \gam_\rad\gg \gam_\s\,.
            \end{array}
            \right.
\end{align}
%%%

(2) Sink is at the center. In this configuration, we choose $z=\ell/2$ for even $\ell$ or $z=(\ell+1)/2$ for odd $\ell$. We approximate the weight by
%%%
\begin{align}   \label{alzmid}
       \al_\ell^{(z)}&\approx\frac{2}{\ell+1}\,.
\end{align}
%%%
This yields the efficiency
%%%
\begin{align}   \label{etazmid}
       \eta_e&\approx \frac{\w_\ell}{\w_0}
        \times \left\{\begin{array}{ccc}
        \ds
            1-\frac{(\ell+1)\gam_\rad}{2\gam_\s}\,,
            & &\ds \frac{\ell+1}{2} \gam_\rad\ll \gam_\s\,,\\\\
        \ds   \frac{2\gam_\s}{(\ell+1)\gam_\rad}\,,
           & &\ds \frac{\ell+1}{2} \gam_\rad\gg \gam_\s\,.
            \end{array}
            \right.
\end{align}
%%%

Notice that \Eqs{etazl} and \eqref{etazmid} contain qubic and linear terms in $\ell$, respectively. This suggests that in longer chain, a sink prepared at the center of the chain is more effective in trapping excitations than a sink positioned at its end. This result will be illustrated in Sec.~\ref{secRad} when we obtain the efficiency numerically.

When the trapping rate of the sink is much greater than the rate of radiation source, the first equation in each of \eqref{etazl} and \eqref{etazmid} has a simple interpretation.
Each excitation introduced by the source acquires energy $\w_0$. For a high power source, the excitation tends to relax to the lowest mode, giving off $2J\cos k_1$ of energy difference to the phonon bath. With a small competition from the radiation loss because of its small rate compared to the absorption rate of the sink, the energy trapped by the sink is then close to $\w_0-2J\cos k_1=\w_\ell$. Hence, the efficiency is $\w_\ell/\w_0$, with a correction term linear in the ratio $\gam_\rad/\gam_\s$.

%%%%%%%%%%%%%%%%%%%%%%%%%%
%       Section         %
%%%%%%%%%%%%%%%%%%%%%%%%%%
\subsection{Low power source}
\label{secLows}

When the energy transfer is sustained by a low power source, we consider $\eta_e$ \eqref{etaEsumsig} in the low and high temperature limit. In the low temperature limit, the excitations will concentrate to the lower energy modes in the steady state under a weak source. The numerator and denominator in $\eta_e$ are then dominated by the $\nb_\ell$ term. Hence, we obtain an expression identical to \Eq{etaElarges}, though under different conditions. This suggests that the behaviour of the energy transfer efficiency under a high power source at moderate temperature is similar to one under a low power source in the low temperature limit. The analysis in Sec.~\ref{secHighs} is then applicable to this situation.

On the other hand, in the high temperature limit the excitations will reach a uniform distribution among all the modes in the steady state. \Eq{etaEsumsig} then yields
%%%
\begin{align}   \label{etaHighT}
       \eta_e&\approx\frac{\ds\ssL\frac{\w_\sig}{\w_0}\al^{(z)}_\sig\gam_\s}{\gam_\s+\ell \gam_\rad}\,,
\end{align}
%%%
where we use $\sum_{\sig=1}^\ell \al_\sig^{(z)}=1$ in the denominator. In the numerator, the second term in the exciton energy $\w_\sig$ \eqref{wb} when multiplied by $\al_\sig^{(z)}$ will give a zero sum over the modes, since we can show that $\sum_\sig \cos k_\sig \sin^2 k_\sig=0$ using the identity in App.~\ref{AppDiscSin}. We then obtain
%%%
\begin{align}   \label{etaHighTb}
       \eta_e&\approx\frac{1}{\ds 1+\frac{\ell \gam_\rad}{\gam_\s}}\,.
\end{align}
%%%
Notice that this expression is independent of the position of the sink. Hence, efficiency is not sensitive to the position of the sink when a weak source is supplied under high temperature bath.

We can consider two opposite limits,
%%%
\begin{align}   \label{etazlows}
       \eta_e&\approx \left\{\begin{array}{ccc}
        \ds
            1-\frac{\ell\gam_\rad}{\gam_\s}\,,
            &\quad &\ds \ell \gam_\rad\ll \gam_\s\,,\\\\
        \ds   \frac{\gam_\s}{\ell\gam_\rad}\,,
           &\quad &\ds \ell\gam_\rad\gg \gam_\s\,.
            \end{array}
            \right.
\end{align}
%%%
When the trapping power of the sink is much larger than the total rate of radiation loss $\ell\gam_\rad$, energy transfer to the sink achieves almost perfect efficiency.

\Eqs{etazl}, \eqref{etazmid} and \eqref{etazlows} also show that efficiency generally decreases as the site number grows bigger.

%%%%%%%%%%%%%%%%%%%%%%%%%%
%        Section         %
%%%%%%%%%%%%%%%%%%%%%%%%%%
\section{Effects of parameters on efficiency}
\label{secEeta}

We use as a reference the typical values of parameters from the well-studied photosynthetic system \cite{Adolphs06,Mohseni08} to evaluate energy transfer efficiency.
For Fenna-Matthews-Olson (FMO) pigment protein complex in green sulphur bacteria, the intersite coupling is approximately $J=100\,\cmi$.
We use $\w_0=12,500 \,\cmi$ as the energy or natural frequency of the sites.
A single site dephases in a phonon bath quite rapidly, with a typical dephasing time of 50 fs. Hence, we use $\gam_\text{d}=20\,\psi$ as a reference, assuming that the dephasing rate is independent of sites.
Exciton can also relax to the ground state through  radiation. The rate $\gam_\rad$ is usually small, with a relaxation time of about 1 ns. Hence, we choose $\gam_\rad=0.001\text{ ps}^{-1}$.
In most studies on the energy transfer efficiency in FMO complex such as in Refs.~\cite{Mohseni08,Olaya08,IshizakiJCP09a,Plenio12}, the system is set off with one excitation that transfers through the complex until it is eventually lost. Using the radiation loss as the longest time scale of the system, we assume that a power of one excitation per nanosecond, or $s=0.001\,\psi$, refers to a low power source.
We choose the inhomogeneity or site-asymmetry to be $\eta=0.1$, and use a trapping power of the sink $\gam_\s=1\,\psi$ as a reference. Finally, we start with a chain without excitation as an initial condition.

The efficiencies of energy transfer plotted in the following graphs are obtained by numerically solving for the roots of the set of rate equations \eqref{dndttot} together with \eqref{ndt} at steady state. The solutions then give the efficiency by means of \Eq{etaEsumsig}.

\begin{figure}[t]
  \centering
\includegraphics[width=3.2in, trim = 5.5cm 11cm 3cm 10cm]{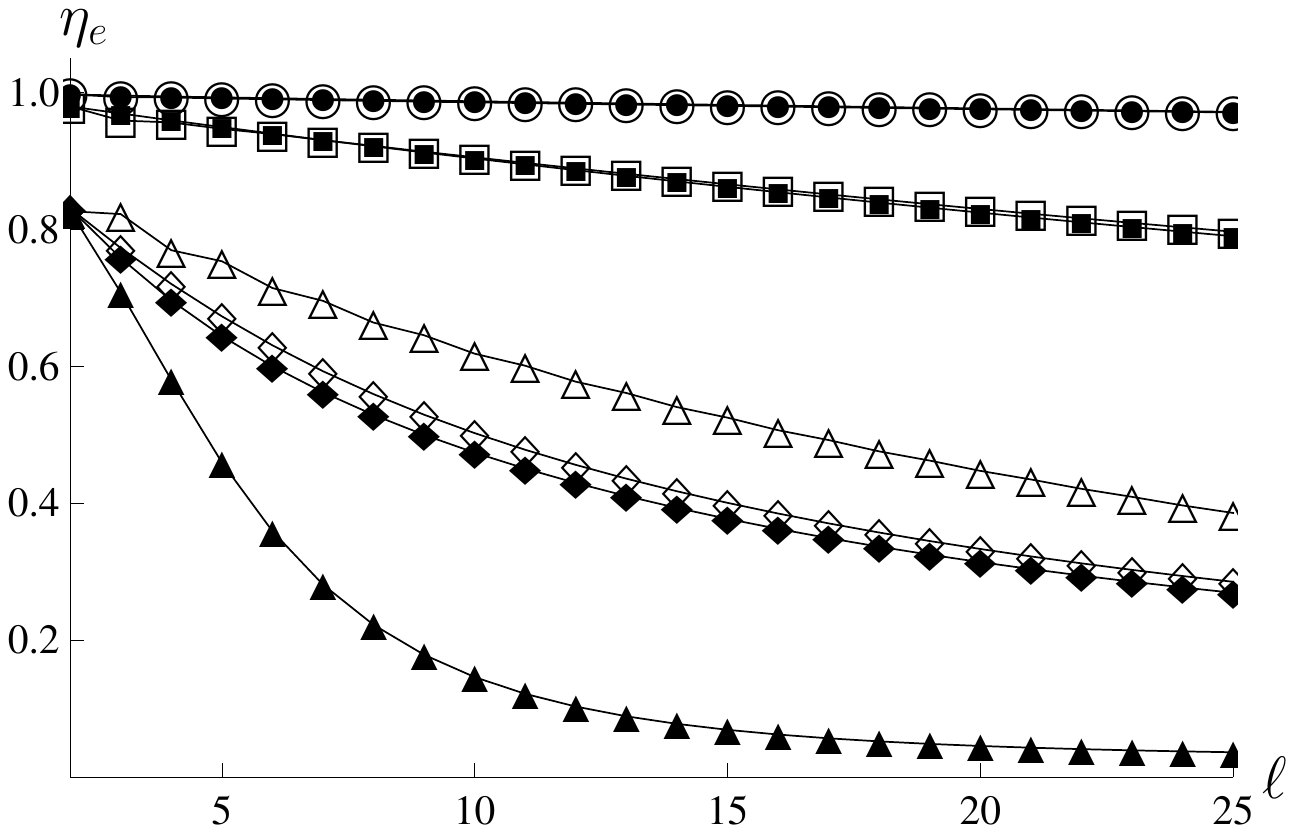}
\caption{Similar symbols denote the same set of parameters. Filled (empty) symbols denote sink at the end (center) of the chain.
In all the curves, $J=100\,\cmi$, $\w_0=12,500\,\cmi$, $T=77$ K, $\eta=0.1$ and $\gam_\text{d}=20\,\psi$. The symbols denote
$\circ=(1, 0.001, 0.1)$,
$\square=(1, 0.01, 0.1)$,
$\lozenge=(0.1, 0.01, 0.1)$, and
$\vartriangle=(0.1, 0.01, 10)$, where
the numbers in brackets refer to $(\gam_\s, \gam_\rad, s)$ in units of $\psi$.
}
\label{fig2}
\end{figure}

We first note that with a temperature of either 77 or 300\,K the above set of reference parameters produces an energy transfer efficiency that is almost perfect, ranging from about 99.8\% for a chain with 2 sites to about 97.2\% for a chain with 25 sites, regardless of whether the sink is placed at the end  of the chain or at its center. Though they are not shown in \Fig{fig2}, the curves of the efficiency for the reference set of parameters almost overlaps with the $\circ$ and $\bullet$ curves in \Fig{fig2}.

%%%
%%%

%%%%%%%%%%%%%%%%%%%%%%%%%%
%        Section         %
%%%%%%%%%%%%%%%%%%%%%%%%%%
\subsection{Parameters external to chain: radiation, source and sink}
\label{secRad}

In \Fig{fig2}, curves labeled by similar shapes refer to the same set of parameters. Sinks located at the end or center of the chain are denoted by filled shapes or empty shapes, respectively.
The temperature of all the curves is fixed at 77 K.
The values of other parameters are listed in \Fig{fig2}.

The first observation we make is that the energy transfer efficiency decreases with an increased number of sites $\ell$.
The loss of energy through radiation and phonon bath are competing with the sink for energy supplied by the source. In the model we consider, there is only one sink available regardless of the number of sites. The power of the sink is distributed across exciton modes according to the weight $\al^{(z)}_\sig$ \eqref{alz}, which becomes even weaker as the number of sites increases.
On the contrary, as the number of sites increases, relatively more energy is prone to loss from the chain since more channels are open to radiation loss and phonon bath.
The total power of loss eventually becomes stronger as the number of sites increases, leading to a greater reduction in the energy transfer efficiency.
%From the point of view of exciton relaxation dynamics, a chain with more sites will have a higher transition rate \eqref{GamlargeL}. When excitations cascade down the energy spectrum faster, energy is lost to phonon bath more rapidly. Furthermore, a redistribution of excitations to lower energy levels also reduces the energy collected by sink. These factors also add up to a reduction in the efficiency.
%The reduction in efficiency becomes more obvious in high power source when relatively more energy is available to radiation loss, as illustrated by the $\blacktriangle$ curve with $s=10\,\psi$, compared to other curves with low power source at $s=0.1\,\psi$ in \Fig{fig2}.

External to the chain related parameters, namely, the radiation rate, the power of sink and source, exert the most significant effects on the energy transfer efficiency at steady state.
For example, a small rate of radiation results in a high efficiency of nearly 98\%, regardless of the position of the sink, as can be seen from the pairs of $(\bullet,\circ)$ curves in \Fig{fig2} which almost overlap among themselves. When we increase the rate of radiation loss from $\gam_\rad=0.001$ to $0.01\,\psi$, the efficiency reduces to about 80\% with a chain of 25 sites, compare the pair of $(\bullet,\circ)$ curves to the pair of $(\blacksquare,\square)$ curves in \Fig{fig2}.

The power of the sink is another main factor that decides the efficiency of energy transfer.
Continue from the set of parameters in the pair of $(\blacksquare,\square)$ curves in \Fig{fig2}, a further decrease in the power of the sink from $\gam_\s=1\,\psi$ to $0.1\,\psi$ causes the efficiency to deteriorate further down to 30\% in a chain with 25 sites, as depicted by the pair of $(\blacklozenge,\lozenge)$ curves in \Fig{fig2}.

The position of the sink could have strong influence on the efficiency at high power source.
When the power of the source increases from $s=0.1\,\psi$ to $10\,\psi$, a sink located at the end of the chain is not so effective in trapping energy from the source compared to a sink placed at its center.
This is illustrated by the $\blacklozenge$ and $\blacktriangle$ curves in \Fig{fig2}, where the efficiency reduces from above 30\% to less than 5\%, respectively, for a chain with 25 sites.

The reverse effect occurs when the sink is positioned at the center of the chain, where an increase in the power of source from $s=0.1$ to $10\,\psi$ produces a rise in efficiency from about 30\% to about 40\% in a chain with 25 sites, compare the $\lozenge$-curve with the $\vartriangle$-curve, respectively.
This indicates that a sink located at the center of the chain is more effective in tapping energy from the chain.

This can be explained by the profile of the excitations in the site basis in \Fig{fig1}.
There, we find that the maximum occupation number occurs at the center of the chain.
Therefore, a sink placed at the center of the chain is more efficient to trap energy from the chain.
The profiles also explain the fact that the position of the sink will not have a significant effect on the efficiency when the excitations distribute uniformly along the chain. This is evident in the pairs of $(\bullet,\circ)$, $(\blacksquare,\square)$ and $(\blacklozenge,\lozenge)$ curves in \Fig{fig2} for a low power source of $s=0.1\,\psi$.

As a high power source of $s=10\,\psi$ is introduced, condensation to the lowest mode occurs. Consequently, the profile of excitation in the $\blacksquare$-curve in \Fig{fig1} has an obvious maximum at the center of the chain. We expect that positing the sink at the its center of the chain can enhance the efficiency greatly. This is evident from \Fig{fig2}, where for a chain of 25 sites, the efficiency increases from below 5\% ($\blacktriangle$ curve) to 40\% ($\vartriangle$ curve) when we reposition the sink from the end of the chain to its center at the same set of parameters. The findings are consistent with the analytic results obtained in Sec.~\ref{secHighs}, compare \Eqs{etazl} and \eqref{etazmid} for sink positioned at the end of the chain and at its center, respectively, where efficiency varies with $\ell^3$ and $\ell$, respectively.

In contrast, the steady state efficiency is not sensitive to the position of the source. Numerical studies show that placing the source at site-1, at the center of the chain, or with its power distributed uniformly over all sites, produce nearly identical efficiency at steady state. We conclude that only the power of the source, not its position, is important in deciding the efficiency at steady state.

i
%%%
\begin{figure}[t]
  \centering
\includegraphics[width=3.2in, trim = 5.5cm 11cm 3cm 10cm]{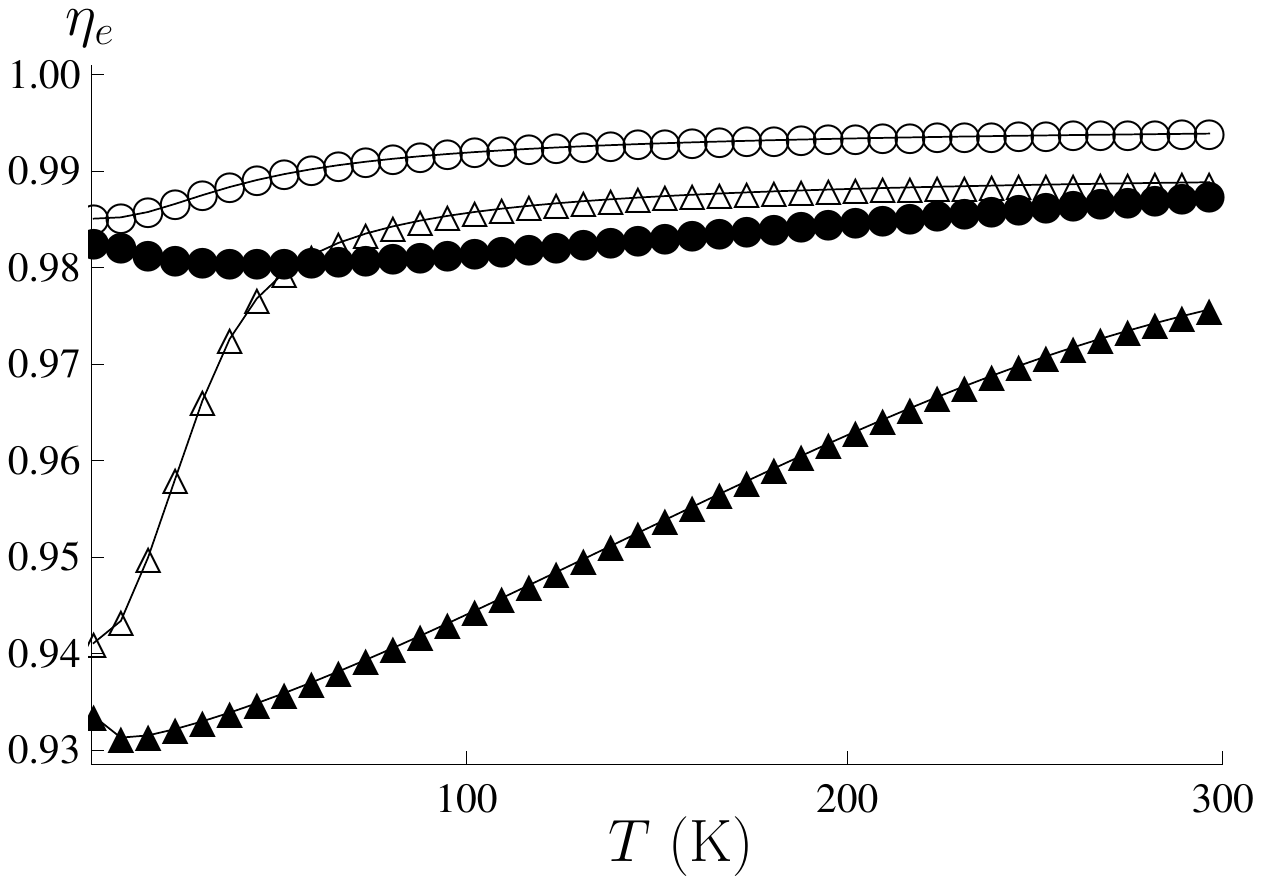}
\caption{In all the curves, $J=100\,\cmi$, $\w_0=12,500\,\cmi$, $\eta=0.1, \gam_\text{d}=20\,\psi, \gam_\s=1\,\psi$ and  $\gam_\rad=0.001\,\psi$. The symbols denote
$\circ=(5,1)$,
$\bullet=(5,10)$,
$\vartriangle=(10,1)$, and
$\blacktriangle=(10,10)$, where the
numbers in brackets denote $(\ell,s\,\psi)$.
}
\label{fig3}
\end{figure}
%%%

%%%%%%%%%%%%%%%%%%%%%%%%%%
%        Section         %
%%%%%%%%%%%%%%%%%%%%%%%%%%
\subsection{Chain related parameters: temperature and intersite coupling}
\label{secT}

In general, we find that an increase of temperature helps in improving efficiency.
In \Fig{fig3}, efficiency is plotted as a function of temperature which ranges from 0 to 300 K for chains with 5 and 10 sites, and with source powers of 1 and $10\,\psi$.
Phonons from higher temperature bath remove excitations trapped in lower levels and redistribute them to higher levels, thus increasing the probability of excitation capture by the sink, whose trapping power distributes among the exciton modes with the factor $\al^{(z)}_\sig$ \eqref{alz}.
We also notice from the pairs of curves in higher power source ($\bullet$ and $\blacktriangle$ curves) that efficiency requires higher temperature to reach its optimum value. High temperature is required because more phonons are required to remove the increase number of excitations trapped in lower energy levels as $s$ increases.

There is a small but interesting effect arises from the number of sites in the chain under a high power source, compare the $\bullet$ and $\blacktriangle$ curves in \Fig{fig3}. Chains with small number of sites ($\ell \lesssim 8$) possess a wide minimum at low temperature between 0 to 100 K in the $\bullet$-curve for $\ell=5$ in \Fig{fig3}. An initial rise of temperature leads to decrease in efficiency, which then increases and eventually reaches its optimum value at higher temperature.
The minimum at low temperature is not prominent in chains with more sites ($\ell \gtrsim 12$), for which rise in temperature always causes the efficiency to increase as explained in the previous paragraph.
This can be seen from the $\blacktriangle$ curve around 0 K in \Fig{fig3}, where the minimum becomes narrower for $\ell=10$ and eventually turns invisible for $\ell\gtrsim12$ (not shown in the figure).

The minimum in the efficiency at low temperature is caused by larger energy gap between transitions in chains with smaller number of sites. We already learned that the width of the energy gap between two exciton levels depends on the factor $\cos k_\mu-\cos k_\nu$, cf.~\Eq{wbmn}, which is wider in chains with less sites. When the bath's temperature is so low that its phonons are not energetic enough to remove excitations out of the low energy level because of the larger gaps, the excitations remain trapped at low levels.
As a result, the efficiency reduces slightly in chains with small number of sites at low temperature when temperature rises slightly. As the number of sites increases, more transitions become available because of the smaller energy gaps.
In this way, removal of excitations trapped at lower energy level becomes plausible even at low temperature. This enhances the probability of excitations trapping by the sink. Efficiency thus increases consistently in longer chain and the minimum at low temperature eventually vanishes as the number of sites increases.

%%%
\begin{figure}[t]
  \centering
\includegraphics[width=3.2in, trim = 5.5cm 11cm 3cm 10cm]{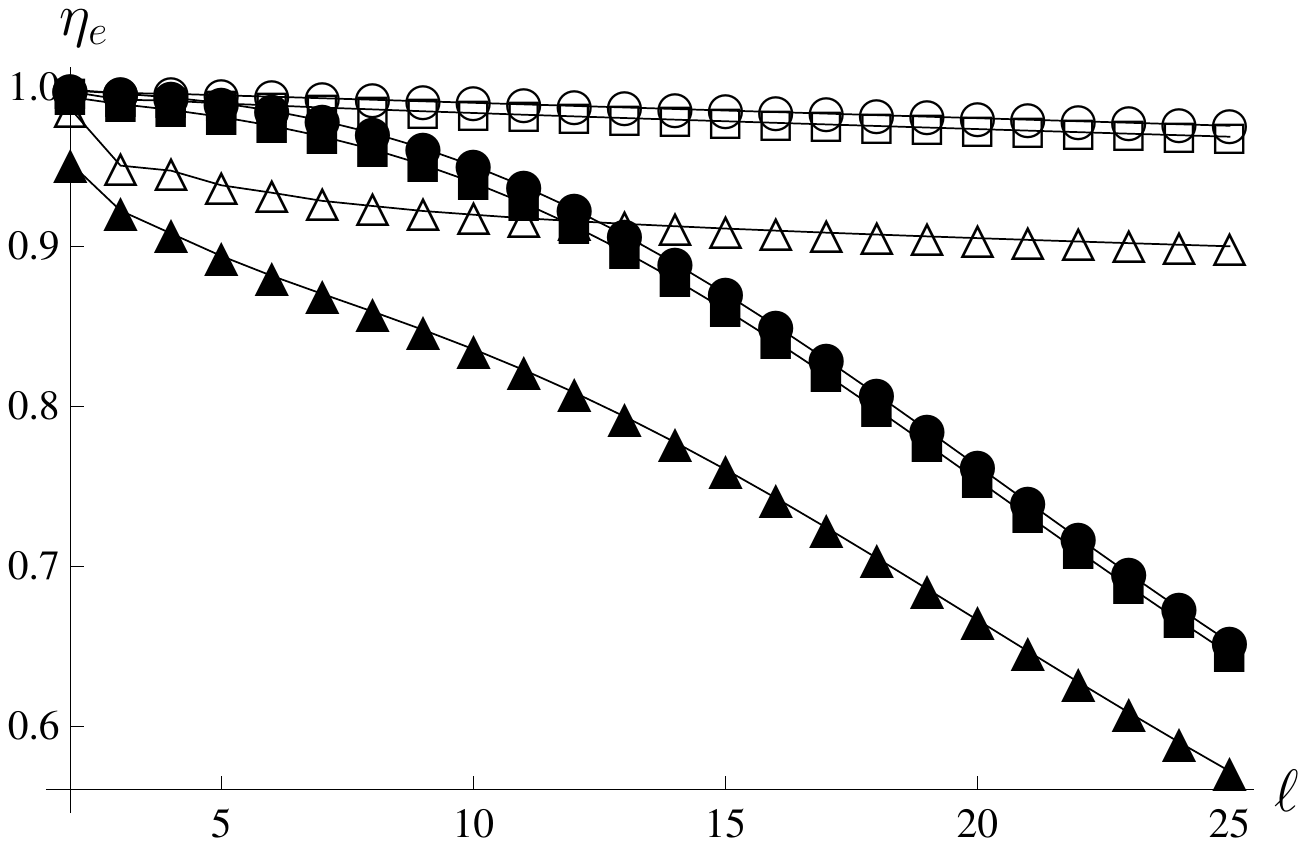}
\caption{
In all the curves, $\w_0=12,500\,\cmi$, $\eta=0.1, \gam_\text{d}=20\,\psi, \gam_\s=1\,\psi$ and $\gam_\rad=0.001\,\psi$.
Numbers in the following braickets refer to $(T\, \text{K}, J \, \cmi, s \, \psi)$. The symbols denote
$\bullet=(7.7, 10, 10^4)$,
$\blacksquare=(77, 100, 10^4)$,
$\blacktriangle=(770, 1000, 10^4)$,
$\circ=(0.77,10,10^3)$,
$\square=(7.7, 100, 10^3)$, and
$\vartriangle=(77, 1000, 10^3)$.
}
\label{fig4}
\end{figure}
%%%

The intersite coupling $J$ affects the relaxation dynamics in that it determines the size of energy gaps \eqref{wbmn}. It scales the temperature by $T'\equiv T/J$, as is seen in the Bose-Einstein distribution \eqref{nbmn} with energy gap \eqref{wbmn}.
Thus, efficiency of curves with similar $T'$ should lie close to each other.
This is shown in the group of $(\bullet,\blacksquare,\blacktriangle)$ and $(\circ,\square,\vartriangle)$ curves in \Fig{fig4}, which have $T'=77$ and 7.7 K$\cdot$cm, respectively.
Since the effect is tiny, we have chosen high power sources to enhance the effect.
The similarity is good for small values of $J=10$ and $100\,\cmi$, where the $(\bullet,\blacksquare)$ and $(\circ,\square)$ curves almost overlap.
However, a larger $J$ results in larger energy gap.
Relaxation of excitations would then cause more energy lost to the phonon bath. This causes the efficiency to reduce relatively more at larger $J=1000\,\cmi$, compare $\blacktriangle$ with the pair ($\bullet, \blacksquare$) curves, and $\vartriangle$ with the pair of $(\circ,\square)$ curves in \Fig{fig4}.

%%%
\begin{figure}[t]
  \centering
\includegraphics[width=3.2in, trim = 5.5cm 11cm 3cm 10cm]{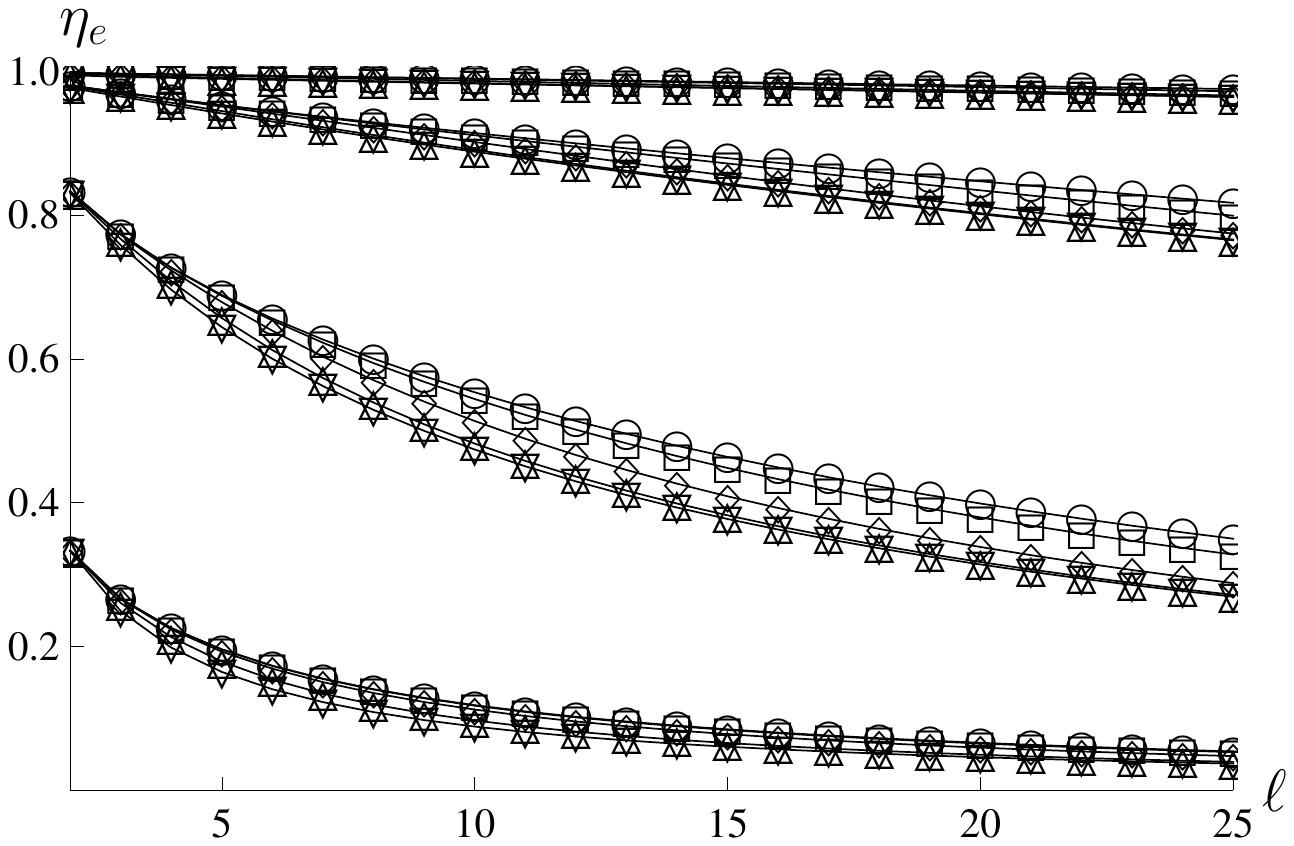}
\caption{In all curves, $T=77$ K, $J=100\,\cmi$, $\w_0=12,500\,\cmi$, $\gam_\s=1\,\psi$, $\eta=0.1, \gam_\text{d}=20\,\psi$ and $s=1\,\psi$. The four groups of curves arranged from the top to the bottom correspond to $\gam_\rad=0.001, 0.01, 0.1, 1\,\psi$, respectively. Within each group, there are five curves labeled by $(\circ,\square,\lozenge,\vartriangle,\triangledown)$ from the top to the bottom, corresponding to $\xi=0.001, 0.01, 0.1, 1, 10$, respectively.
}
\label{fig5}
\end{figure}
%%%

%%%%%%%%%%%%%%%%%%%%%%%%%%
%        Section         %
%%%%%%%%%%%%%%%%%%%%%%%%%%
\subsection{Model related parameters: inhomogeneity and dephasing rate}
\label{secet}

From the expression of $\Gam_\mn$ \eqref{Gam} and $\gam^{(2)}_\mn$ \eqref{gam2}, we notice that $\eta$ and $\gam_\text{d}$ influence the reduced dynamics in the form $\eta^2\gam_\text{d}$. We study their effects by introducing a numerical factor $\xi$ that scales them as $\xi\eta^2\gam_\text{d}$.
Four groups of curves are plotted in \Fig{fig5}.
Arranged from the top to the bottom, they correspond to $\gam_\rad=0.001, 0.01, 0.1$ and $1\,\psi$, respectively.
Within each group, there are five curves arranged from the top to the bottom, denoted by $(\circ,\square,\lozenge,\vartriangle,\triangledown)$, respectively. They respectively correspond to the values of $\xi$ as it varies from 0.001 to 10 in multiples of 10. We find that the efficiency gradually reduces with the increase of $\xi$.
Greater value of $\xi$ yields higher transition rate $\Gam_\mn$.
More rapid relaxation rate then competes with the trapping power of the sink to induce more loss of energy to phonon bath, leading to smaller efficiency.

This is a small effect. From \Fig{fig5}, we find that a change of $\xi$ across four orders of magnitude alters the efficiency by less than 10\% for curves with intermediate efficiency. The change in efficiency is less than 3\% when the curves are in both extremes of its efficiency, cf.~the highest and the lowest group of curves in \Fig{fig5}. In fact, they almost overlap in these cases. Therefore, the efficiency obtained is quite robust against the change of the inhomogeneity $\eta$, and the dephasing rate $\gam_\text{d}$.
In \Fig{fig5} we have chosen a temperature of 77 K. The change in the efficiency is even smaller at 300 K.

%%%
\begin{figure}[t]
  \centering
\includegraphics[width=3.2in, trim = 5.5cm 11cm 3cm 10cm]{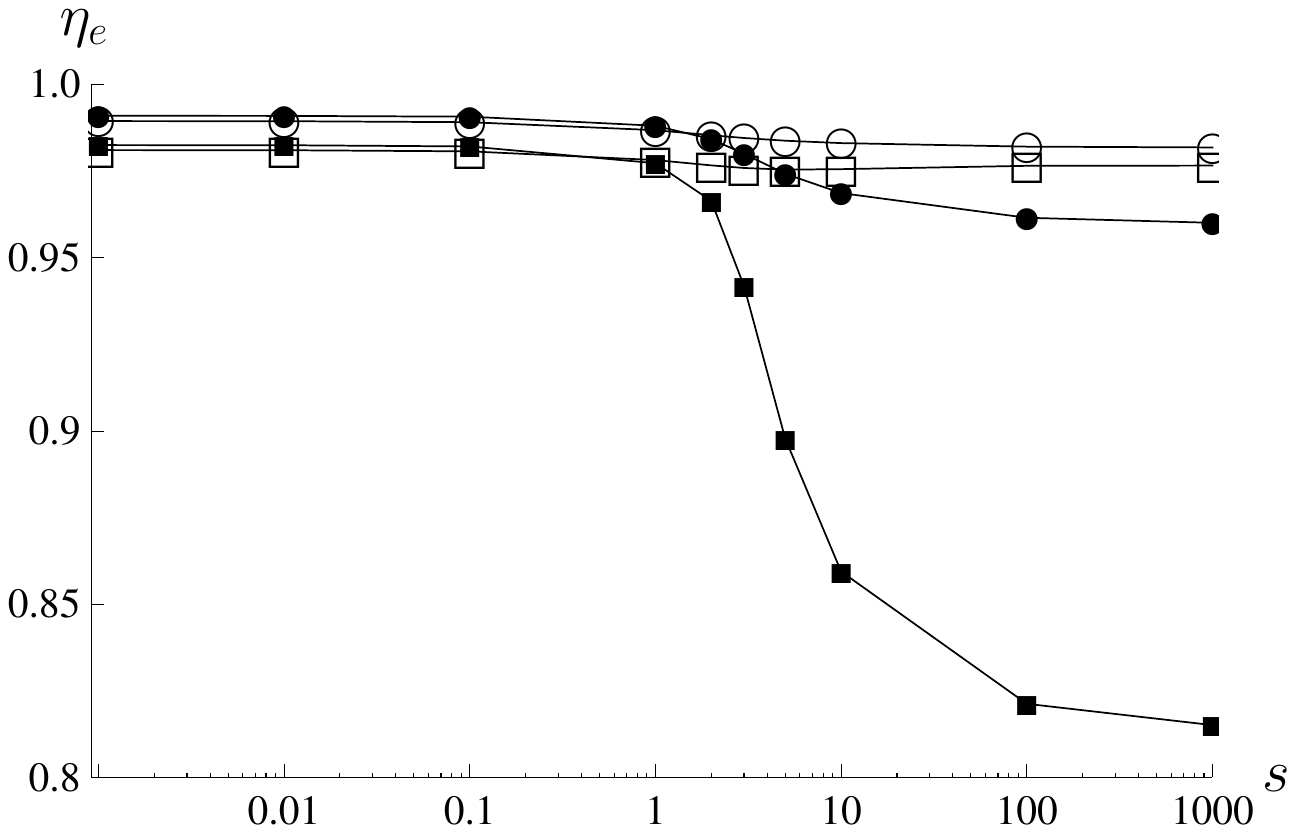}
\caption{In all curves, $T=77$ K, $J=100\,\cmi$, $\w_0=12,500\,\cmi$, $\eta=0.1, \gam_\text{d}=20\,\psi$, $\gam_\s=1\,\psi$  and $\gam_\text{d}=0.001\,\psi$. Source power $s$ is in units of $\psi$, plotted in natural logarithmic scale. Circles and squares denote chains with $\ell=7$ and 15, respectively. Filled and empty shapes label configurations with sinks prepared at the end of the chain and at its center, respectively.
}
\label{fig6}
\end{figure}
%%%

%%%%%%%%%%%%%%%%%%%%%%%%%%
%        Section         %
%%%%%%%%%%%%%%%%%%%%%%%%%%
\subsection{Change in efficiency during condensation}
\label{secetas}

We learn in Sec.~\ref{secSrc} that as the source power becomes sufficiently high, excitations will condense to the lowest energy level. Let us consider the behaviour of energy transfer efficiency during condensation using the reference set of parameters. \Fig{fig6} plots efficiency against source power in natural logarithmic scale for chains with $\ell=7$ and 15, denoted by circles and squares, respectively. Configurations with a sink prepared at the end of the chain and at its center are labelled by filled and empty shapes, respectively. The power increases from 0.001 to $10^3$ $\psi$.

We notice that under a weak source, efficiency is not sensitive to the power and the position of the sink, though the efficiency is slightly better with a sink prepared at the end of the chain. As the source power turns into the stronger region, efficiency for configuration with sink prepared at the center decreases slightly. However, for sink prepared at the end of the chain, the decrease is abrupt when excitations start to condense to the lowest energy level between $s=1$ to 10 $\psi$. This change is more apparent in longer chain. We learn from \Fig{fig1} that the profile of excitation in the site basis has a prominent maximum at the center of the chain when excitations concentrate to the lowest energy level. Therefore, a sink prepared at the end of the chain is not as effective as a sink placed at its center in trapping energy from the chain. The efficiency gradually stabilizes at greater source power.

%%%%%%%%%%%%%%%%%%%%%%%%%%
%        Section         %
%%%%%%%%%%%%%%%%%%%%%%%%%%
\section{Conclusion}

We have studied the reduced dynamics of molecular chains coupled weakly to phonon bath with small inhomogeneity. The excitation relaxation dynamics is largely determined by the transition rate between modes lying next to each other in the excitation energy spectrum. Due to collective effect, the coherence lifetime between different excitation modes in a chain is longer compared to single site dephasing time.
However, as the length of the chain increases, a rapid rise in the transition rate eventually reverses the effect.

The rapid rise in the exciton relaxation rates with the length of chain leads to a clear separation of the time scale in the system into a shorter one determined by the exciton transition rate, and a longer one dictated by external influence such as the rate of radiation loss, and the trapping rate and position of the sink.
As a consequence, the efficiency of energy transfer at steady state is not sensitive to the change of internal parameters related to the chain and phonon. Parameters external to the chain and phonon, such as the pumping rate of the source, the rate of radiation loss, and the trapping rate and the position of the sink, play a much bigger role in deciding the efficiency of energy transfer.

We learn that higher correlation functions give rise to nonlinear terms in the rate equations of the excitation occupation number, which influence the distribution of the excitations at steady state. Their effects are most prominent under strong source, when they cause the concentration of excitations to the lowest energy mode. They reveal themselves when a sink prepared at the center of the chain is more efficient in trapping energy than a sink placed at its end.

Even though in this work we use the parameters from a specific photosynthetic system for illustrations, our analysis on the excitation relaxation dynamics and steady state energy transfer is general and hence is relevant to the transport of energy in other systems.

\acknowledgments
Support by the Ministry of Higher Education, Malaysia (MOHE) under the Fundamental Research Grant Scheme (FRGS), Grant No.~FRGS/1/2020/STG07/UNIM/02/1, is gratefully acknowledged.

\appendix

\begin{widetext}

%%%%%%%%%%%%%%%%%%%%%%%%%%
%       Section         %
%%%%%%%%%%%%%%%%%%%%%%%%%%
\section{Discrete sine transform}
\label{AppDiscSin}

A finite sum of exponential functions gives
%%%
%\begin{widetext}
\begin{align}   \label{Prope}
    \sum_{x=1}^{\ell+1} e^{i k_\mu x} &= \left\{ \begin{array}{cllllll}
    \ell+1, &\quad &\mu=0,2m(\ell+1),&  & m=1,2,\cdots, \\
    0, &\quad &\mu=2,4,6,\cdots,  & \text{excluding } 2m(\ell+1), & m=1,2,\cdots,    \\
%    -1,    &\mu=(2m+1)(\ell+1),  &\text{for odd } \ell+1,& m=0,1,2,\cdots, \\
    -1+i\cot\left(\frac{1}{2}k_\mu\right), &\quad &\mu=1,3,5,\cdots.
    \end{array}
    \right.
\end{align}
%%%
The real and imaginary parts of \Eq{Prope} are
%%%
\begin{align}     \label{Propcos}
    \sum_{x=1}^{\ell+1} \cos(k_\mu x)&=\left\{ \begin{array}{cllllll}
    \ell+1,&\quad  &\mu=0, 2m(\ell+1),  &\quad &m=1,2,\cdots, \\
    0, &\quad &\mu=2,4,6,\cdots, & \text{excluding } 2m(\ell+1),  &m=1,2,\cdots,   \\
    -1,  &\quad  &\mu=1,3,5,\cdots,
    \end{array}
    \right.
\end{align}
%%%
\begin{align}       \label{Propsin}
    \sum_{x=1}^{\ell+1} \sin(k_\mu x)&=\left\{ \begin{array}{cll}
    0,&\quad &\mu=0,2,4,\cdots,  \\
%    0,&\mu=(2m+1)(\ell+1),  &\text{for odd } \ell+1, & m=0,1,2,\cdots,  \\
    \cot\left(\frac{1}{2}k_\mu\right), &\quad &\mu=1,3,5,\cdots.
    \end{array}
    \right.
\end{align}
\end{widetext}
%%%
Using the identities, we obtain the inverse of \Eq{A},
%%%
\begin{align}   \label{ax}
    a_x&=\sqrt{\frac{2}{\ell+1}}\smL \sin(k_\mu x) \, A_\mu \,.
\end{align}
%%%

In the exciton basis, the first term of $H'_0$ \eqref{H0'} becomes
%%%
\begin{align}   \label{H0App}
        \siL\w'_x \adg_x a_x &=\frac{2}{\ell+1}\sum_{\mu,\nu=1}^{\ell}  \Adg_\mu A_\nu \nm
        &\quad\times\left(\siL \w'_x\sin( k_\mu x)\sin (k_\nu x)\right)\,.
\end{align}
%%%
Since $\w'_x$ \eqref{wx'} depends on the site index $x$, the sum over $x$ cannot be carried out explicitly. To overcome this problem, we assume that the correction term to $\w_0$ in \Eq{wx'} is negligible, and approximate $\w'_x$ by $\w_0$. We can then carried out the sum in the bracket to obtain
%%%
\begin{align}   \label{Snn}
        \siL \sin( k_\mu x)\sin (k_\nu x)=\frac{1}{2} (\ell+1)\del_\mn  \,,
\end{align}
%%%
where $\del_\mn $ is the kronecker-delta function.
Substituting \Eq{Snn} into the approximate expression of \Eq{H0App} then yields the diagonalized form of the first term of $H'_0$ \eqref{H0'},
%%%
\begin{align}
        \siL\w'_x \adg_x a_x &\approx \sum_{\mu=1}^{\ell}\w_0   \Adg_\mu A_\mu\,.
\end{align}
%%%

Substituting $a_x$ \eqref{ax} and its hermitian conjugate into the second term of $H'_0$ \eqref{H0'} we obtain
%%%
\begin{align}   \label{H0App2}
    &J \siLm \big(\adg_x a_{x+1}+a_x\adg_{x+1}\big)\nm
    &\quad=\frac{2J}{\ell+1}\left( \smL h_{\mu\mu} \Adg_\mu A_\mu+  \sideset{}{'}\sum_{\mu,\nu=1}^\ell h_\mn  \Adg_\mu A_\nu \right)\,,
\end{align}
%%%
where $\sum'_{\mu,\nu=1}$ denotes a sum over $\mu$ and $\nu$ but excluding $\mu=\nu$.
The coefficient $h_\mn$ is
%%%
\begin{align}   \label{Smn}
        h_\mn &\equiv \siL \big[ \sin\big( k_\mu x\big)\sin \big(k_\nu (x+1)\big) \nm
        &\qquad\qquad+ \sin(\big( k_\mu (x+1)\big)\sin \big(k_\nu x\big) \big]\,.
\end{align}
%%%
This sum can be evaluated to give
%%%
\begin{align}
       h_\mn &=(\ell+1)\cos(k_\mu) \del_\mn  \,.
\end{align}
%%%
Substituting it into \Eq{H0App2} gives
%%%
\begin{align}   \label{H0AppJ}
    J \siLm \big(\adg_x a_{x+1}+a_x\adg_{x+1}\big)%\nm
    &=2J \smL\cos(k_\mu) \Adg_\mu A_\mu\,.
\end{align}
%%%
As a result, the discrete sine transform diagonalizes $H'_0$ \eqref{H0'} into \Eq{H0A}, with energy $\w_\mu$ given by \Eq{wb}.

\begin{widetext}

%%%%%%%%%%%%%%%%%%%%%%%%%%
%       Section         %
%%%%%%%%%%%%%%%%%%%%%%%%%%
\section{Full set of rate equations}
\label{AppCompRedDyn}

The complete set of coupled nonlinear rate equations inclusive of the correlation components is
%%%
\begin{align}   \label{ndtcompi}
    \frac{dn_\sig}{dt}\bigg|_\ch&=
    \sum_{\mu<\sig} \Gam_\ms \big[n_\mu(n^\th_\ms+1)+n_\sig(n_\mu-n^\th_\ms)
    +|n_\ms|^2+|m_\ms|^2\big]\nm
    &\quad-\sum_{\nu>\sig} \Gam_\sn\big[n_\sig(n^\th_\sn +1)+n_\nu(n_\sig-n^\th_\sn)
    +|n_\sn|^2+|m_\sn|^2\big]\,,
\end{align}
%%%
%%%
\begin{align}   \label{nskidt}
    \frac{dn_\sk}{dt}\bigg|_\ch&=-i(\w_\sig-\w_\kap)n_\sk \nm
    &-\half\left[\sum_{\mu<\sig} \Gam_\ms n^\th_\ms +\sum_{\mu<\kap} \Gam_\mk n^\th_\mk\right] n_\sk
    +\half\left[\sum_{\mu<\sig} \Gam_\ms +\sum_{\mu<\kap} \Gam_\mk\right]
    \big(n_\mu n_\sk+n^*_\ms n_\mk+m^*_\ms m_\mk\big)\nm
    &-\half\left[\sum_{\nu>\sig}\Gam_\sn (n^\th_\sn+1)+\sum_{\nu>\kap}\Gam_\kn(n^\th_\kn+1)\right]n_\sk -\half\left[\sum_{\nu>\sig}\Gam_\sn+\sum_{\nu>\kap}\Gam_\kn\right]
    \big(n_\sk n_\nu+n_\sn n^*_\kn+m^*_\sn m_\kn\big)\,,\nm
    &\qquad
    \sig<\kap\,,
\end{align}
%%%
%%%
\begin{align}   \label{mskdt}
    \frac{dm_\sk}{dt}\bigg|_\ch&=i(\w_\sig+\w_\kap)m_\sk\nm
    &-\half\left[\sum_{\mu<\sig} \Gam_\ms n^\th_\ms +\sum_{\mu<\kap} \Gam_\mk n^\th_\mk\right] m_\sk
    +\half\left[\sum_{\mu<\sig} \Gam_\ms+\sum_{\mu<\kap} \Gam_\mk\right]
    \big(n_\mu m_\sk+n_\ms m_\mk+m_\ms n_\mk \big)\nm
    &-\half\left[\sum_{\nu>\sig}\Gam_\sn (n^\th_\sn+1)+\sum_{\nu>\kap}\Gam_\kn(n^\th_\kn+1)\right]m_\sk -\half\left[\sum_{\nu>\sig}\Gam_\sn+\sum_{\nu>\kap}\Gam_\kn\right]
    \big(n_\nu m_\sk+n^*_\sn m_\kn+m_\sn n^*_\kn \big)\,,\nm
    &\qquad \sig\leq \kap\,.
\end{align}
%%%

\end{widetext}

%%%%%%%%%%%%%%%%%%%%%%%%%%
%       Section         %
%%%%%%%%%%%%%%%%%%%%%%%%%%
\section{External source and sink}
\label{AppExt}

We can model external source and sink connected to the chain at site-$z$ by coupling the oscillator's operator to the respective field through the interaction,
%%%
\begin{align}   \label{sourceH}
    H_\ext=\lam \sum_k v_k( a_z \cdg_k+ \adg_z c_k)\,,
\end{align}
%%%
where $\lam$ denotes a dimensionless coupling constant, $v_k$ is a real form factor, and $c^\dg_k, c_k$ are the creation and annihilation operators of the field mode.
In the exciton basis, $H_\ext$ becomes
%%%
\begin{align}   \label{sourceHexc}
    H_\ext=\lam \smL \sum_k \al^{(z)}_\mu v_k( A_\mu \cdg_k+ \Adg_\mu c_k)\,,
\end{align}
%%%
where $\al^{(z)}_\mu$ is a weight factor defined in \Eq{alz}.
$H_\ext$ gives rise to a dissipative reduced dynamics $\d\rho/\d t|_\ext=-K_\ext \rho$, where
%%%
\begin{align}   \label{Ksource}
    K_\ext\rho&=-\frac{1}{2}\ssL \al^{(z)}_\sig\gam^\ext_\sig\nm
    &\times\bigg[ (n^\th_\sig+1)
         (2A_\sig\rho \Adg_\sig -\Adg_\sig A_\sig \rho-\rho \Adg_\sig A_\sig)\nm
    &\qquad+n^\th_\sig
         (2\Adg_\sig\rho A_\sig -A_\sig\Adg_\sig \rho-\rho A_\sig\Adg_\sig )\bigg]\,,\\
        n^\th_\sig&\equiv \frac{1}{e^{\bt\w_\sig}-1}\,,
\end{align}
%%%
with the relaxation rate
%%%
\begin{align} \label{eps}
    \gam^\ext_\sig&\equiv 2\pi\lam^2\sum_k v_k^2\del(\w_k-\w_\sig)\,.
\end{align}
%%%
It yields the following rate equation of exciton occupation number of the $\sig$-mode,
%%%
\begin{align}   \label{sknT}
        \frac{dn_\sig}{dt}\bigg|_\ext&=\al^{(z)}_\sig\big(\Phi_\sig-\gam^\ext_\sig n_\sig\big)\,,\\
        \Phi_\sig&\equiv \gam^\ext_\sig n^\th_\sig\,,
\end{align}
%%%
where $\Phi_\sig$ functions like a source term.

To describe a ``pure" source connected to the chain at site $z=1$, we assume that $\Phi_\sig=s$, where $s$ is the number of excitations per unit time fed by the source to the chain, is independent of exciton modes. It should be much greater than the relaxation term $\gam^\ext_\sig n_\sig$ in \Eq{sknT} to give
%%%
\begin{align}   \label{extSrc}
        \frac{dn_\sig}{dt}\bigg|_\src&=\al^{(1)}_\sig s\,.
\end{align}
%%%
We also assume that the source does not create new correlations in the chain.

A ``pure" sink is obtained by taking the bath's temperature to be zero, $n^\th_\sig=0$.
As a result, we are led to consider the following time evolution of the exciton occupation number
%%%
\begin{align}   \label{extSk}
        \frac{dn_\sig}{dt}\bigg|_\si&=-\al^{(z)}_\sig\gam_\s n_\sig\,,
\end{align}
%%%
where $\gam_\s$ is the trapping rate of the sink.

When a ``pure" sink is connected to site-$z$ of the chain, the following terms are added to the time evolution of the correlation components,
%%%
\begin{align}   \label{Snmn}
       \frac{dn_\sk}{dt}\bigg|_\si&=-\half
       \big(\al^{(z)}_\sig
       +\al^{(z)}_\kap\big)\gam_\s n_\sk\,,\\
        \frac{dm_\sk}{dt}\bigg|_\si&=-\half
        \big(\al^{(z)}_\sig
        +\al^{(z)}_\kap\big)\gam_\s m_\sk\,.\label{Smmn}
\end{align}
%%%

%%%%%%%%%%%%%%%%%%%%%%%%%%
%       Section         %
%%%%%%%%%%%%%%%%%%%%%%%%%%
\section{Efficiency based on time evolution}
\label{Apptranseta}

In this appendix we show that the energy transfer efficiency obtained by considering the time evolution of the system \cite{Mohseni08,Olaya08} approaches the efficiency at steady state \eqref{etaEsumsig} in the long time limit.

The total energy absorbed by the sink up to a time $\tau$ is
%%%
\begin{align}   \label{esk}
        E_\si(\tau)&=-\int_{0}^{\tau} \ssL\w_\sig \frac{dn_\sig}{dt}\bigg|_\si dt
        &=\ssL\w_\sig\al^{(z)}_\sig\gam_\s a_\sig(\tau)\,,
\end{align}
%%%
where
%%%
\begin{align}   \label{atau}
        a_\sig(\tau)&\equiv\int_{0}^{\tau} n_\sig(t)dt
\end{align}
%%%
is the area enclosed by the curve $n_\sig(t)$ and the time axis up to time $\tau$.
The total energy supplied by the source up to time $\tau$ is
%%%
\begin{align}   \label{esrc}
        E_\src(\tau)&=\int_{0}^{\tau}\w_0 s \,dt
        =\w_0\ssL\eps^{(z)}_\sig A_\sig(\tau)\,,
\end{align}
%%%
where we substitute $s$ using \Eq{sumnt} to get the third equality, in which
%%%
\begin{align}   \label{Anb}
        A_\sig(\tau)&\equiv \int_{0}^{\tau} \nb_\sig \,dt =\nb_\sig\tau
\end{align}
%%%
is the area enclosed by the steady state occupation number $\nb_\sig=n_\sig(t\rightarrow\infty)$ and the time axis up to time $\tau$.

Since $n_\sig(t)$ would have reached the steady state after some finite time, the ratio $(A_\sig-a_\sig)/A_\sig\rightarrow0$ as $\tau\rightarrow\infty$.
Hence, we expect that in the long time limit $a_\sig/A_\sig$ should approach 1,
%%%
\begin{align}   \label{aA1}
       \frac{a_\sig(\tau)}{A_\sig(\tau)}\xrightarrow{\tau\rightarrow\infty} 1\,.
\end{align}
%%%

The energy transfer efficiency up to time $\tau$ is defined as
%%%
\begin{align}   \label{etae}
        \eta_E(\tau)&\equiv\frac{E_\si(\tau)}{E_\src(\tau)}\,.
\end{align}
%%%
It can be cast into the form
%%%
\begin{align}   \label{etaeaA}
        \eta_E(\tau)&=\frac{\ds\ssL\frac{\w_\sig}{\w_0}\gam_\s\al^{(z)}_\sig A_\sig(\tau) \cdot \frac{a_\sig(\tau)}{A_\sig(\tau)}}{\ds\ssL\eps^{(z)}_\sig A_\sig(\tau)}\,.
\end{align}
%%%
It approaches $\eta_e$ \eqref{etaEsumsig} in the long $\tau$ limit
%%%
\begin{align}
        \eta_E(\tau)&\xrightarrow{\tau\rightarrow \infty}
        \eta_e\,,
\end{align}
%%%
after using \Eqs{aA1} and \eqref{Anb}.

%\bibliography{SSchain}{}

\begin{thebibliography}{38}%
\makeatletter
\providecommand \@ifxundefined [1]{%
 \@ifx{#1\undefined}
}%
\providecommand \@ifnum [1]{%
 \ifnum #1\expandafter \@firstoftwo
 \else \expandafter \@secondoftwo
 \fi
}%
\providecommand \@ifx [1]{%
 \ifx #1\expandafter \@firstoftwo
 \else \expandafter \@secondoftwo
 \fi
}%
\providecommand \natexlab [1]{#1}%
\providecommand \enquote  [1]{``#1''}%
\providecommand \bibnamefont  [1]{#1}%
\providecommand \bibfnamefont [1]{#1}%
\providecommand \citenamefont [1]{#1}%
\providecommand \href@noop [0]{\@secondoftwo}%
\providecommand \href [0]{\begingroup \@sanitize@url \@href}%
\providecommand \@href[1]{\@@startlink{#1}\@@href}%
\providecommand \@@href[1]{\endgroup#1\@@endlink}%
\providecommand \@sanitize@url [0]{\catcode `\\12\catcode `\$12\catcode
  `\&12\catcode `\#12\catcode `\^12\catcode `\_12\catcode `\%12\relax}%
\providecommand \@@startlink[1]{}%
\providecommand \@@endlink[0]{}%
\providecommand \url  [0]{\begingroup\@sanitize@url \@url }%
\providecommand \@url [1]{\endgroup\@href {#1}{\urlprefix }}%
\providecommand \urlprefix  [0]{URL }%
\providecommand \Eprint [0]{\href }%
\providecommand \doibase [0]{https://doi.org/}%
\providecommand \selectlanguage [0]{\@gobble}%
\providecommand \bibinfo  [0]{\@secondoftwo}%
\providecommand \bibfield  [0]{\@secondoftwo}%
\providecommand \translation [1]{[#1]}%
\providecommand \BibitemOpen [0]{}%
\providecommand \bibitemStop [0]{}%
\providecommand \bibitemNoStop [0]{.\EOS\space}%
\providecommand \EOS [0]{\spacefactor3000\relax}%
\providecommand \BibitemShut  [1]{\csname bibitem#1\endcsname}%
\let\auto@bib@innerbib\@empty
%</preamble>
\bibitem [{\citenamefont {Breuer}\ and\ \citenamefont
  {Petruccione}(2002)}]{Breuer}%
  \BibitemOpen
  \bibfield  {author} {\bibinfo {author} {\bibfnamefont {H.-P.}\ \bibnamefont
  {Breuer}}\ and\ \bibinfo {author} {\bibfnamefont {F.}~\bibnamefont
  {Petruccione}},\ }\href@noop {} {\emph {\bibinfo {title} {The Theory of Open
  Quantum Systems}}}\ (\bibinfo  {publisher} {Oxford University Press},\
  \bibinfo {address} {Oxford},\ \bibinfo {year} {2002})\BibitemShut {NoStop}%
\bibitem [{\citenamefont {Kenkre}\ and\ \citenamefont
  {Reineker}(1982)}]{Kenkre82}%
  \BibitemOpen
  \bibinfo {editor} {\bibfnamefont {V.~M.}\ \bibnamefont {Kenkre}}\ and\
  \bibinfo {editor} {\bibfnamefont {P.}~\bibnamefont {Reineker}},\ eds.,\
  \href@noop {} {\emph {\bibinfo {title} {Exciton Dynamics in Molecular
  Crystals and Aggregates}}}\ (\bibinfo  {publisher} {Springer},\ \bibinfo
  {address} {Berlin},\ \bibinfo {year} {1982})\BibitemShut {NoStop}%
\bibitem [{\citenamefont {May}\ and\ \citenamefont {K\"{u}hn}(2011)}]{May11}%
  \BibitemOpen
  \bibfield  {author} {\bibinfo {author} {\bibfnamefont {V.}~\bibnamefont
  {May}}\ and\ \bibinfo {author} {\bibfnamefont {O.}~\bibnamefont {K\"{u}hn}},\
  }\href@noop {} {\emph {\bibinfo {title} {Charge and Energy Transfer Dynamics
  in Molecular Systems}}},\ \bibinfo {edition} {3rd}\ ed.\ (\bibinfo
  {publisher} {Wiley-VCH},\ \bibinfo {address} {Weinheim},\ \bibinfo {year}
  {2011})\BibitemShut {NoStop}%
\bibitem [{\citenamefont {Jang}(2020)}]{Jang20}%
  \BibitemOpen
  \bibinfo {editor} {\bibfnamefont {S.~J.}\ \bibnamefont {Jang}},\ ed.,\ \href
  {https://doi.org/https://doi.org/10.1016/B978-0-08-102335-8.00005-3} {\emph
  {\bibinfo {title} {Dynamics of Molecular Excitons}}},\ Nanophotonics\
  (\bibinfo  {publisher} {Elsevier},\ \bibinfo {year} {2020})\BibitemShut
  {NoStop}%
\bibitem [{\citenamefont {Leegwater}\ \emph {et~al.}(1997)\citenamefont
  {Leegwater}, \citenamefont {Durrant},\ and\ \citenamefont
  {Klug}}]{Leegwater97}%
  \BibitemOpen
  \bibfield  {author} {\bibinfo {author} {\bibfnamefont {J.~A.}\ \bibnamefont
  {Leegwater}}, \bibinfo {author} {\bibfnamefont {J.~R.}\ \bibnamefont
  {Durrant}},\ and\ \bibinfo {author} {\bibfnamefont {D.~R.}\ \bibnamefont
  {Klug}},\ }\bibfield  {title} {\bibinfo {title} {Exciton equilibration
  induced by phonons:  theory and application to ps ii reaction centers},\
  }\href {https://doi.org/10.1021/jp9634058} {\bibfield  {journal} {\bibinfo
  {journal} {J. Phys. Chem. B}\ }\textbf {\bibinfo {volume} {101}},\ \bibinfo
  {pages} {7205} (\bibinfo {year} {1997})}\BibitemShut {NoStop}%
\bibitem [{\citenamefont {Adolphs}\ and\ \citenamefont
  {Renger}(2006)}]{Adolphs06}%
  \BibitemOpen
  \bibfield  {author} {\bibinfo {author} {\bibfnamefont {J.}~\bibnamefont
  {Adolphs}}\ and\ \bibinfo {author} {\bibfnamefont {T.}~\bibnamefont
  {Renger}},\ }\bibfield  {title} {\bibinfo {title} {How proteins trigger
  excitation energy transfer in the fmo complex of green sulfur bacteria},\
  }\href {https://doi.org/10.1529/biophysj.105.079483} {\bibfield  {journal}
  {\bibinfo  {journal} {Biophys. J.}\ }\textbf {\bibinfo {volume} {91}},\
  \bibinfo {pages} {2778} (\bibinfo {year} {2006})}\BibitemShut {NoStop}%
\bibitem [{\citenamefont {Mohseni}\ \emph {et~al.}(2008)\citenamefont
  {Mohseni}, \citenamefont {Rebentrost}, \citenamefont {Lloyd},\ and\
  \citenamefont {Aspuru-Guzik}}]{Mohseni08}%
  \BibitemOpen
  \bibfield  {author} {\bibinfo {author} {\bibfnamefont {M.}~\bibnamefont
  {Mohseni}}, \bibinfo {author} {\bibfnamefont {P.}~\bibnamefont {Rebentrost}},
  \bibinfo {author} {\bibfnamefont {S.}~\bibnamefont {Lloyd}},\ and\ \bibinfo
  {author} {\bibfnamefont {A.}~\bibnamefont {Aspuru-Guzik}},\ }\bibfield
  {title} {\bibinfo {title} {Environment-assisted quantum walks in
  photosynthetic energy transfer},\ }\href {https://doi.org/10.1063/1.3002335}
  {\bibfield  {journal} {\bibinfo  {journal} {J. Chem. Phys.}\ }\textbf
  {\bibinfo {volume} {129}},\ \bibinfo {pages} {174106} (\bibinfo {year}
  {2008})}\BibitemShut {NoStop}%
\bibitem [{\citenamefont {Olaya-Castro}\ \emph {et~al.}(2008)\citenamefont
  {Olaya-Castro}, \citenamefont {Lee}, \citenamefont {Olsen},\ and\
  \citenamefont {Johnson}}]{Olaya08}%
  \BibitemOpen
  \bibfield  {author} {\bibinfo {author} {\bibfnamefont {A.}~\bibnamefont
  {Olaya-Castro}}, \bibinfo {author} {\bibfnamefont {C.~F.}\ \bibnamefont
  {Lee}}, \bibinfo {author} {\bibfnamefont {F.~F.}\ \bibnamefont {Olsen}},\
  and\ \bibinfo {author} {\bibfnamefont {N.~F.}\ \bibnamefont {Johnson}},\
  }\bibfield  {title} {\bibinfo {title} {Efficiency of energy transfer in a
  light-harvesting system under quantum coherence},\ }\href
  {https://doi.org/10.1103/PhysRevB.78.085115} {\bibfield  {journal} {\bibinfo
  {journal} {Phys. Rev. B}\ }\textbf {\bibinfo {volume} {78}},\ \bibinfo
  {pages} {085115} (\bibinfo {year} {2008})}\BibitemShut {NoStop}%
\bibitem [{\citenamefont {Ishizaki}\ and\ \citenamefont
  {Fleming}(2009{\natexlab{a}})}]{IshizakiPNAS09}%
  \BibitemOpen
  \bibfield  {author} {\bibinfo {author} {\bibfnamefont {A.}~\bibnamefont
  {Ishizaki}}\ and\ \bibinfo {author} {\bibfnamefont {G.~R.}\ \bibnamefont
  {Fleming}},\ }\bibfield  {title} {\bibinfo {title} {Theoretical examination
  of quantum coherence in a photosynthetic system at physiological
  temperature},\ }\href {https://doi.org/10.1073/pnas.0908989106} {\bibfield
  {journal} {\bibinfo  {journal} {Proc. Natl. Acad. Sci. USA}\ }\textbf
  {\bibinfo {volume} {106}},\ \bibinfo {pages} {17255} (\bibinfo {year}
  {2009}{\natexlab{a}})}\BibitemShut {NoStop}%
\bibitem [{\citenamefont {Ishizaki}\ and\ \citenamefont
  {Fleming}(2009{\natexlab{b}})}]{IshizakiJCP09a}%
  \BibitemOpen
  \bibfield  {author} {\bibinfo {author} {\bibfnamefont {A.}~\bibnamefont
  {Ishizaki}}\ and\ \bibinfo {author} {\bibfnamefont {G.~R.}\ \bibnamefont
  {Fleming}},\ }\bibfield  {title} {\bibinfo {title} {On the adequacy of the
  redfield equation and related approaches to the study of quantum dynamics in
  electronic energy transfer},\ }\href
  {https://doi.org/http://dx.doi.org/10.1063/1.3155214} {\bibfield  {journal}
  {\bibinfo  {journal} {J. Chem. Phys.}\ }\textbf {\bibinfo {volume} {130}},\
  \bibinfo {eid} {234111} (\bibinfo {year} {2009}{\natexlab{b}})}\BibitemShut
  {NoStop}%
\bibitem [{\citenamefont {Jang}\ and\ \citenamefont {Mennucci}(2018)}]{Jang18}%
  \BibitemOpen
  \bibfield  {author} {\bibinfo {author} {\bibfnamefont {S.~J.}\ \bibnamefont
  {Jang}}\ and\ \bibinfo {author} {\bibfnamefont {B.}~\bibnamefont
  {Mennucci}},\ }\bibfield  {title} {\bibinfo {title} {Delocalized excitons in
  natural light-harvesting complexes},\ }\href
  {https://doi.org/10.1103/RevModPhys.90.035003} {\bibfield  {journal}
  {\bibinfo  {journal} {Rev. Mod. Phys.}\ }\textbf {\bibinfo {volume} {90}},\
  \bibinfo {pages} {035003} (\bibinfo {year} {2018})}\BibitemShut {NoStop}%
\bibitem [{\citenamefont {Davydov}(1979)}]{Davydov79}%
  \BibitemOpen
  \bibfield  {author} {\bibinfo {author} {\bibfnamefont {A.~S.}\ \bibnamefont
  {Davydov}},\ }\bibfield  {title} {\bibinfo {title} {Solitons in molecular
  systems},\ }\href {http://iopscience.iop.org/1402-4896/20/3-4/013} {\bibfield
   {journal} {\bibinfo  {journal} {Phys. Scr.}\ }\textbf {\bibinfo {volume}
  {20}},\ \bibinfo {pages} {387} (\bibinfo {year} {1979})}\BibitemShut
  {NoStop}%
\bibitem [{\citenamefont {Christiansen}\ and\ \citenamefont
  {Scott}(1990)}]{Davydov90}%
  \BibitemOpen
  \bibinfo {editor} {\bibfnamefont {P.~L.}\ \bibnamefont {Christiansen}}\ and\
  \bibinfo {editor} {\bibfnamefont {A.~C.}\ \bibnamefont {Scott}},\ eds.,\
  \href@noop {} {\emph {\bibinfo {title} {Davydov's Soliton Revisited:
  Self-Trapping of Vibrational Energy in Protein}}}\ (\bibinfo  {publisher}
  {Springer},\ \bibinfo {address} {New York},\ \bibinfo {year}
  {1990})\BibitemShut {NoStop}%
\bibitem [{\citenamefont {Coropceanu}\ \emph {et~al.}(2007)\citenamefont
  {Coropceanu}, \citenamefont {Cornil}, \citenamefont {da~Silva~Filho},
  \citenamefont {Olivier}, \citenamefont {Silbey},\ and\ \citenamefont
  {Br\'{e}das}}]{Coropceanu07}%
  \BibitemOpen
  \bibfield  {author} {\bibinfo {author} {\bibfnamefont {V.}~\bibnamefont
  {Coropceanu}}, \bibinfo {author} {\bibfnamefont {J.}~\bibnamefont {Cornil}},
  \bibinfo {author} {\bibfnamefont {D.~A.}\ \bibnamefont {da~Silva~Filho}},
  \bibinfo {author} {\bibfnamefont {Y.}~\bibnamefont {Olivier}}, \bibinfo
  {author} {\bibfnamefont {R.}~\bibnamefont {Silbey}},\ and\ \bibinfo {author}
  {\bibfnamefont {J.-L.}\ \bibnamefont {Br\'{e}das}},\ }\bibfield  {title}
  {\bibinfo {title} {Charge transport in organic semiconductors},\ }\href
  {https://doi.org/10.1021/cr050140x} {\bibfield  {journal} {\bibinfo
  {journal} {Chem. Rev.}\ }\textbf {\bibinfo {volume} {107}},\ \bibinfo {pages}
  {926} (\bibinfo {year} {2007})}\BibitemShut {NoStop}%
\bibitem [{\citenamefont {Maier}\ \emph {et~al.}(2019)\citenamefont {Maier},
  \citenamefont {Brydges}, \citenamefont {Jurcevic}, \citenamefont {Trautmann},
  \citenamefont {Hempel}, \citenamefont {Lanyon}, \citenamefont {Hauke},
  \citenamefont {Blatt},\ and\ \citenamefont {Roos}}]{Maier19}%
  \BibitemOpen
  \bibfield  {author} {\bibinfo {author} {\bibfnamefont {C.}~\bibnamefont
  {Maier}}, \bibinfo {author} {\bibfnamefont {T.}~\bibnamefont {Brydges}},
  \bibinfo {author} {\bibfnamefont {P.}~\bibnamefont {Jurcevic}}, \bibinfo
  {author} {\bibfnamefont {N.}~\bibnamefont {Trautmann}}, \bibinfo {author}
  {\bibfnamefont {C.}~\bibnamefont {Hempel}}, \bibinfo {author} {\bibfnamefont
  {B.~P.}\ \bibnamefont {Lanyon}}, \bibinfo {author} {\bibfnamefont
  {P.}~\bibnamefont {Hauke}}, \bibinfo {author} {\bibfnamefont
  {R.}~\bibnamefont {Blatt}},\ and\ \bibinfo {author} {\bibfnamefont {C.~F.}\
  \bibnamefont {Roos}},\ }\bibfield  {title} {\bibinfo {title}
  {Environment-assisted quantum transport in a 10-qubit network},\ }\href
  {https://doi.org/10.1103/PhysRevLett.122.050501} {\bibfield  {journal}
  {\bibinfo  {journal} {Phys. Rev. Lett.}\ }\textbf {\bibinfo {volume} {122}},\
  \bibinfo {pages} {050501} (\bibinfo {year} {2019})}\BibitemShut {NoStop}%
\bibitem [{\citenamefont {Plenio}\ and\ \citenamefont
  {Huelga}(2008)}]{Plenio08}%
  \BibitemOpen
  \bibfield  {author} {\bibinfo {author} {\bibfnamefont {M.~B.}\ \bibnamefont
  {Plenio}}\ and\ \bibinfo {author} {\bibfnamefont {S.~F.}\ \bibnamefont
  {Huelga}},\ }\bibfield  {title} {\bibinfo {title} {Dephasing-assisted
  transport: quantum networks and biomolecules},\ }\href
  {https://doi.org/10.1088/1367-2630/10/11/113019} {\bibfield  {journal}
  {\bibinfo  {journal} {New J. Phys.}\ }\textbf {\bibinfo {volume} {10}},\
  \bibinfo {pages} {113019} (\bibinfo {year} {2008})}\BibitemShut {NoStop}%
\bibitem [{\citenamefont {Chin}\ \emph {et~al.}(2010)\citenamefont {Chin},
  \citenamefont {Datta}, \citenamefont {Caruso}, \citenamefont {Huelga},\ and\
  \citenamefont {Plenio}}]{Chin10}%
  \BibitemOpen
  \bibfield  {author} {\bibinfo {author} {\bibfnamefont {A.~W.}\ \bibnamefont
  {Chin}}, \bibinfo {author} {\bibfnamefont {A.}~\bibnamefont {Datta}},
  \bibinfo {author} {\bibfnamefont {F.}~\bibnamefont {Caruso}}, \bibinfo
  {author} {\bibfnamefont {S.~F.}\ \bibnamefont {Huelga}},\ and\ \bibinfo
  {author} {\bibfnamefont {M.~B.}\ \bibnamefont {Plenio}},\ }\bibfield  {title}
  {\bibinfo {title} {Noise-assisted energy transfer in quantum networks and
  light-harvesting complexes},\ }\href
  {https://doi.org/10.1088/1367-2630/12/6/065002} {\bibfield  {journal}
  {\bibinfo  {journal} {New J. Phys.}\ }\textbf {\bibinfo {volume} {12}},\
  \bibinfo {pages} {065002} (\bibinfo {year} {2010})}\BibitemShut {NoStop}%
i\bibitem [{\citenamefont {Chin}\ \emph {et~al.}(2012)\citenamefont {Chin},
  \citenamefont {Huelga},\ and\ \citenamefont {Plenio}}]{Plenio12}%
  \BibitemOpen
  \bibfield  {author} {\bibinfo {author} {\bibfnamefont {A.~W.}\ \bibnamefont
  {Chin}}, \bibinfo {author} {\bibfnamefont {S.~F.}\ \bibnamefont {Huelga}},\
  and\ \bibinfo {author} {\bibfnamefont {M.~B.}\ \bibnamefont {Plenio}},\
  }\bibfield  {title} {\bibinfo {title} {Coherence and decoherence in
  biological systems: principles of noise-assisted transport and the origin of
  long-lived coherences},\ }\href {https://doi.org/10.1098/rsta.2011.0224}
  {\bibfield  {journal} {\bibinfo  {journal} {Philos. Trans. R. Soc., A}\
  }\textbf {\bibinfo {volume} {370}},\ \bibinfo {pages} {3638} (\bibinfo {year}
  {2012})}\BibitemShut {NoStop}%
\bibitem [{\citenamefont {Cao}\ and\ \citenamefont {Silbey}(2009)}]{Cao09}%
  \BibitemOpen
  \bibfield  {author} {\bibinfo {author} {\bibfnamefont {J.}~\bibnamefont
  {Cao}}\ and\ \bibinfo {author} {\bibfnamefont {R.~J.}\ \bibnamefont
  {Silbey}},\ }\bibfield  {title} {\bibinfo {title} {Optimization of exciton
  trapping in energy transfer processes},\ }\href
  {https://doi.org/10.1021/jp9032589} {\bibfield  {journal} {\bibinfo
  {journal} {J. Phys. Chem. A}\ }\textbf {\bibinfo {volume} {113}},\ \bibinfo
  {pages} {13825} (\bibinfo {year} {2009})}\BibitemShut {NoStop}%
\bibitem [{\citenamefont {Yang}\ and\ \citenamefont {Cao}(2020)}]{Cao20}%
  \BibitemOpen
  \bibfield  {author} {\bibinfo {author} {\bibfnamefont {P.-Y.}~\bibnamefont
  {Yang}}\ and\ \bibinfo {author} {\bibfnamefont {J.}\ \bibnamefont
  {Cao}},\ }\bibfield  {title} {\bibinfo {title} {Steady-State Analysis of Light-Harvesting Energy Transfer Driven by Incoherent Light: From Dimers to Networks},\ }\href
  {https://doi.org/10.1021/acs.jpclett.0c01648} {\bibfield  {journal} {\bibinfo
  {journal} {J. Phys. Chem. Lett.}\ }\textbf {\bibinfo {volume} {11}},\ \bibinfo
  {pages} {7204} (\bibinfo {year} {2020}{\natexlab{b}})}\BibitemShut {NoStop}%i
\bibitem [{\citenamefont {Engel}\ \emph {et~al.}(2007)\citenamefont {Engel},
  \citenamefont {Calhoun}, \citenamefont {Read}, \citenamefont {Ahn},
  \citenamefont {Mancal}, \citenamefont {Cheng}, \citenamefont {Blankenship},\
  and\ \citenamefont {Fleming}}]{Engel07}%
  \BibitemOpen
  \bibfield  {author} {\bibinfo {author} {\bibfnamefont {G.~S.}\ \bibnamefont
  {Engel}}, \bibinfo {author} {\bibfnamefont {T.~R.}\ \bibnamefont {Calhoun}},
  \bibinfo {author} {\bibfnamefont {E.~L.}\ \bibnamefont {Read}}, \bibinfo
  {author} {\bibfnamefont {T.-K.}\ \bibnamefont {Ahn}}, \bibinfo {author}
  {\bibfnamefont {T.}~\bibnamefont {Man\v{c}al}}, \bibinfo {author} {\bibfnamefont
  {Y.-C.}\ \bibnamefont {Cheng}}, \bibinfo {author} {\bibfnamefont {R.~E.}\
  \bibnamefont {Blankenship}},\ and\ \bibinfo {author} {\bibfnamefont {G.~R.}\
  \bibnamefont {Fleming}},\ }\bibfield  {title} {\bibinfo {title} {Evidence for
  wavelike energy transfer through quantum coherence in photosynthetic
  systems},\ }\href {https://doi.org/10.1038/nature05678} {\bibfield  {journal}
  {\bibinfo  {journal} {Nature (London)}\ }\textbf {\bibinfo {volume} {446}},\
  \bibinfo {pages} {782} (\bibinfo {year} {2007})}\BibitemShut {NoStop}%
\bibitem [{\citenamefont {Panitchayangkoon}\ \emph {et~al.}(2010)\citenamefont
  {Panitchayangkoon}, \citenamefont {Hayes}, \citenamefont {Fransted},
  \citenamefont {Caram}, \citenamefont {Harel}, \citenamefont {Wen},
  \citenamefont {Blankenship},\ and\ \citenamefont
  {Engel}}]{Panitchayangkoon10}%
  \BibitemOpen
  \bibfield  {author} {\bibinfo {author} {\bibfnamefont {G.}~\bibnamefont
  {Panitchayangkoon}}, \bibinfo {author} {\bibfnamefont {D.}~\bibnamefont
  {Hayes}}, \bibinfo {author} {\bibfnamefont {K.~A.}\ \bibnamefont {Fransted}},
  \bibinfo {author} {\bibfnamefont {J.~R.}\ \bibnamefont {Caram}}, \bibinfo
  {author} {\bibfnamefont {E.}~\bibnamefont {Harel}}, \bibinfo {author}
  {\bibfnamefont {J.}~\bibnamefont {Wen}}, \bibinfo {author} {\bibfnamefont
  {R.~E.}\ \bibnamefont {Blankenship}},\ and\ \bibinfo {author} {\bibfnamefont
  {G.~S.}\ \bibnamefont {Engel}},\ }\bibfield  {title} {\bibinfo {title}
  {Long-lived quantum coherence in photosynthetic complexes at physiological
  temperature},\ }\href {https://doi.org/10.1073/pnas.1005484107} {\bibfield
  {journal} {\bibinfo  {journal} {Proc. Natl. Acad. Sci. USA}\ }\textbf
  {\bibinfo {volume} {107}},\ \bibinfo {pages} {12766} (\bibinfo {year}
  {2010})}\BibitemShut {NoStop}%
\bibitem [{\citenamefont {Kassal}\ \emph {et~al.}(2013)\citenamefont {Kassal},
  \citenamefont {Yuen-Zhou},\ and\ \citenamefont {Rahimi-Keshari}}]{Kassal13}%
  \BibitemOpen
  \bibfield  {author} {\bibinfo {author} {\bibfnamefont {I.}~\bibnamefont
  {Kassal}}, \bibinfo {author} {\bibfnamefont {J.}~\bibnamefont {Yuen-Zhou}},\
  and\ \bibinfo {author} {\bibfnamefont {S.}~\bibnamefont {Rahimi-Keshari}},\
  }\bibfield  {title} {\bibinfo {title} {Does coherence enhance transport in
  photosynthesis?},\ }\href {https://doi.org/10.1021/jz301872b} {\bibfield
  {journal} {\bibinfo  {journal} {J. Phys. Chem. Lett.}\ }\textbf {\bibinfo
  {volume} {4}},\ \bibinfo {pages} {362} (\bibinfo {year} {2013})}\BibitemShut
  {NoStop}%
\bibitem [{\citenamefont {Duan}\ \emph {et~al.}(2017)\citenamefont {Duan},
  \citenamefont {Prokhorenko}, \citenamefont {Cogdell}, \citenamefont {Ashraf},
  \citenamefont {Stevens}, \citenamefont {Thorwart},\ and\ \citenamefont
  {Miller}}]{Duan17}%
  \BibitemOpen
  \bibfield  {author} {\bibinfo {author} {\bibfnamefont {H.-G.}\ \bibnamefont
  {Duan}}, \bibinfo {author} {\bibfnamefont {V.~I.}\ \bibnamefont
  {Prokhorenko}}, \bibinfo {author} {\bibfnamefont {R.~J.}\ \bibnamefont
  {Cogdell}}, \bibinfo {author} {\bibfnamefont {K.}~\bibnamefont {Ashraf}},
  \bibinfo {author} {\bibfnamefont {A.~L.}\ \bibnamefont {Stevens}}, \bibinfo
  {author} {\bibfnamefont {M.}~\bibnamefont {Thorwart}},\ and\ \bibinfo
  {author} {\bibfnamefont {R.~J.~D.}\ \bibnamefont {Miller}},\ }\bibfield
  {title} {\bibinfo {title} {Nature does not rely on long-lived electronic
  quantum coherence for photosynthetic energy transfer},\ }\href
  {https://doi.org/10.1073/pnas.1702261114} {\bibfield  {journal} {\bibinfo
  {journal} {Proc. Natl. Acad. Sci. USA}\ }\textbf {\bibinfo {volume} {114}},\
  \bibinfo {pages} {8493} (\bibinfo {year} {2017})}\BibitemShut {NoStop}%
\bibitem [{\citenamefont {Pach\'{o}n}\ and\ \citenamefont
  {Brumer}(2011)}]{Pachon11}%
  \BibitemOpen
  \bibfield  {author} {\bibinfo {author} {\bibfnamefont {L.~A.}\ \bibnamefont
  {Pach\'{o}n}}\ and\ \bibinfo {author} {\bibfnamefont {P.}~\bibnamefont
  {Brumer}},\ }\bibfield  {title} {\bibinfo {title} {Physical basis for
  long-lived electronic coherence in photosynthetic light-harvesting systems},\
  }\href {https://doi.org/10.1021/jz201189p} {\bibfield  {journal} {\bibinfo
  {journal} {J. Phys. Chem. Lett.}\ }\textbf {\bibinfo {volume} {2}},\ \bibinfo
  {pages} {2728} (\bibinfo {year} {2011})}\BibitemShut {NoStop}%
\bibitem [{\citenamefont {Christensson}\ \emph {et~al.}(2012)\citenamefont
  {Christensson}, \citenamefont {Kauffmann}, \citenamefont {Pullerits},\ and\
  \citenamefont {Mančal}}]{Christensson12}%
  \BibitemOpen
  \bibfield  {author} {\bibinfo {author} {\bibfnamefont {N.}~\bibnamefont
  {Christensson}}, \bibinfo {author} {\bibfnamefont {H.~F.}\ \bibnamefont
  {Kauffmann}}, \bibinfo {author} {\bibfnamefont {T.}~\bibnamefont
  {Pullerits}},\ and\ \bibinfo {author} {\bibfnamefont {T.}~\bibnamefont
  {Man\v{c}al}},\ }\bibfield  {title} {\bibinfo {title} {Origin of long-lived
  coherences in light-harvesting complexes},\ }\href
  {https://doi.org/10.1021/jp304649c} {\bibfield  {journal} {\bibinfo
  {journal} {J. Phys. Chem. B}\ }\textbf {\bibinfo {volume} {116}},\ \bibinfo
  {pages} {7449} (\bibinfo {year} {2012})}\BibitemShut {NoStop}%
\bibitem [{\citenamefont {Fr\"ohlich}(1968{\natexlab{a}})}]{Frohlich68}%
  \BibitemOpen
  \bibfield  {author} {\bibinfo {author} {\bibfnamefont {H.}~\bibnamefont
  {Fr\"ohlich}},\ }\bibfield  {title} {\bibinfo {title} {Bose condensation of
  strongly excited longitudinal electric modes},\ }\href
  {https://doi.org/https://doi.org/10.1016/0375-9601(68)90242-9} {\bibfield
  {journal} {\bibinfo  {journal} {Phys. Lett. A}\ }\textbf {\bibinfo {volume}
  {26}},\ \bibinfo {pages} {402 } (\bibinfo {year}
  {1968}{\natexlab{a}})}\BibitemShut {NoStop}%
\bibitem [{\citenamefont {Fr\"ohlich}(1968{\natexlab{b}})}]{Frohlich68b}%
  \BibitemOpen
  \bibfield  {author} {\bibinfo {author} {\bibfnamefont {H.}~\bibnamefont
  {Fr\"ohlich}},\ }\bibfield  {title} {\bibinfo {title} {Long-range coherence
  and energy storage in biological systems},\ }\href
  {https://doi.org/10.1002/qua.560020505} {\bibfield  {journal} {\bibinfo
  {journal} {Int. J. Quant. Chem.}\ }\textbf {\bibinfo {volume} {2}},\ \bibinfo
  {pages} {641} (\bibinfo {year} {1968}{\natexlab{b}})}\BibitemShut {NoStop}%
\bibitem [{\citenamefont {Mills}(1983)}]{Mills83}%
  \BibitemOpen
  \bibfield  {author} {\bibinfo {author} {\bibfnamefont {R.~E.}\ \bibnamefont
  {Mills}},\ }\bibfield  {title} {\bibinfo {title} {Fr\"ohlich's model of
  nonthermal excitations in biological systems},\ }\href
  {https://doi.org/10.1103/PhysRevA.28.379} {\bibfield  {journal} {\bibinfo
  {journal} {Phys. Rev. A}\ }\textbf {\bibinfo {volume} {28}},\ \bibinfo
  {pages} {379} (\bibinfo {year} {1983})}\BibitemShut {NoStop}%
\bibitem [{\citenamefont {Jesenko}\ and\ \citenamefont
  {\v{Z}nidari\v{c}}(2013)}]{Jesenko13}%
  \BibitemOpen
  \bibfield  {author} {\bibinfo {author} {\bibfnamefont {S.}~\bibnamefont
  {Jesenko}}\ and\ \bibinfo {author} {\bibfnamefont {M.}~\bibnamefont
  {\v{Z}nidari\v{c}}},\ }\bibfield  {title} {\bibinfo {title} {Excitation
  energy transfer efficiency: Equivalence of transient and stationary setting
  and the absence of non-markovian effects},\ }\href
  {https://doi.org/10.1063/1.4802816} {\bibfield  {journal} {\bibinfo
  {journal} {J. Chem. Phys.}\ }\textbf {\bibinfo {volume} {138}},\ \bibinfo
  {pages} {174103} (\bibinfo {year} {2013})}\BibitemShut {NoStop}%
\bibitem [{\citenamefont {Tay}(2014)}]{Tay14}%
  \BibitemOpen
  \bibfield  {author} {\bibinfo {author} {\bibfnamefont {B.~A.}\ \bibnamefont
  {Tay}},\ }\bibfield  {title} {\bibinfo {title} {Attenuation of excitation
  decay rate due to collective effect},\ }\href
  {https://doi.org/10.1103/PhysRevE.90.022142} {\bibfield  {journal} {\bibinfo
  {journal} {Phys. Rev. E}\ }\textbf {\bibinfo {volume} {90}},\ \bibinfo
  {pages} {022142} (\bibinfo {year} {2014})}\BibitemShut {NoStop}%
\bibitem [{\citenamefont {Yarkony}\ and\ \citenamefont
  {Silbey}(1976)}]{Yarkony76}%
  \BibitemOpen
  \bibfield  {author} {\bibinfo {author} {\bibfnamefont {D.}~\bibnamefont
  {Yarkony}}\ and\ \bibinfo {author} {\bibfnamefont {R.}~\bibnamefont
  {Silbey}},\ }\bibfield  {title} {\bibinfo {title} {Comments on exciton phonon
  coupling: Temperature dependence},\ }\href
  {https://doi.org/http://dx.doi.org/10.1063/1.433182} {\bibfield  {journal}
  {\bibinfo  {journal} {J. Chem. Phys.}\ }\textbf {\bibinfo {volume} {65}},\
  \bibinfo {pages} {1042} (\bibinfo {year} {1976})}\BibitemShut {NoStop}%
\bibitem [{\citenamefont {Brown}\ \emph {et~al.}(1986)\citenamefont {Brown},
  \citenamefont {Lindenberg},\ and\ \citenamefont {West}}]{Brown86}%
  \BibitemOpen
  \bibfield  {author} {\bibinfo {author} {\bibfnamefont {D.~W.}\ \bibnamefont
  {Brown}}, \bibinfo {author} {\bibfnamefont {K.}~\bibnamefont {Lindenberg}},\
  and\ \bibinfo {author} {\bibfnamefont {B.~J.}\ \bibnamefont {West}},\
  }\bibfield  {title} {\bibinfo {title} {On the dynamics of polaron formation
  in a deformable medium},\ }\href
  {https://doi.org/http://dx.doi.org/10.1063/1.450502} {\bibfield  {journal}
  {\bibinfo  {journal} {J. Chem. Phys.}\ }\textbf {\bibinfo {volume} {84}},\
  \bibinfo {pages} {1574} (\bibinfo {year} {1986})}\BibitemShut {NoStop}%
\bibitem [{\citenamefont {Nardecchia}\ \emph {et~al.}(2018)\citenamefont
  {Nardecchia}, \citenamefont {Torres}, \citenamefont {Lechelon}, \citenamefont
  {Giliberti}, \citenamefont {Ortolani}, \citenamefont {Nouvel}, \citenamefont
  {Gori}, \citenamefont {Meriguet}, \citenamefont {Donato}, \citenamefont
  {Preto}, \citenamefont {Varani}, \citenamefont {Sturgis},\ and\ \citenamefont
  {Pettini}}]{Nardecchia18}%
  \BibitemOpen
  \bibfield  {author} {\bibinfo {author} {\bibfnamefont {I.}~\bibnamefont
  {Nardecchia}}, \bibinfo {author} {\bibfnamefont {J.}~\bibnamefont {Torres}},
  \bibinfo {author} {\bibfnamefont {M.}~\bibnamefont {Lechelon}}, \bibinfo
  {author} {\bibfnamefont {V.}~\bibnamefont {Giliberti}}, \bibinfo {author}
  {\bibfnamefont {M.}~\bibnamefont {Ortolani}}, \bibinfo {author}
  {\bibfnamefont {P.}~\bibnamefont {Nouvel}}, \bibinfo {author} {\bibfnamefont
  {M.}~\bibnamefont {Gori}}, \bibinfo {author} {\bibfnamefont {Y.}~\bibnamefont
  {Meriguet}}, \bibinfo {author} {\bibfnamefont {I.}~\bibnamefont {Donato}},
  \bibinfo {author} {\bibfnamefont {J.}~\bibnamefont {Preto}}, \bibinfo
  {author} {\bibfnamefont {L.}~\bibnamefont {Varani}}, \bibinfo {author}
  {\bibfnamefont {J.}~\bibnamefont {Sturgis}},\ and\ \bibinfo {author}
  {\bibfnamefont {M.}~\bibnamefont {Pettini}},\ }\bibfield  {title} {\bibinfo
  {title} {Out-of-equilibrium collective oscillation as phonon condensation in
  a model protein},\ }\href {https://doi.org/10.1103/PhysRevX.8.031061}
  {\bibfield  {journal} {\bibinfo  {journal} {Phys. Rev. X}\ }\textbf {\bibinfo
  {volume} {8}},\ \bibinfo {pages} {031061} (\bibinfo {year}
  {2018})}\BibitemShut {NoStop}%
\bibitem [{\citenamefont {Tay}(2013)}]{Tay13}%
  \BibitemOpen
  \bibfield  {author} {\bibinfo {author} {\bibfnamefont {B.~A.}\ \bibnamefont
  {Tay}},\ }\bibfield  {title} {\bibinfo {title} {Reduced dynamics of two
  oscillators collectively coupled to a thermal bath},\ }\href
  {https://doi.org/10.1103/PhysRevE.87.052117} {\bibfield  {journal} {\bibinfo
  {journal} {Phys. Rev. E}\ }\textbf {\bibinfo {volume} {87}},\ \bibinfo
  {pages} {052117} (\bibinfo {year} {2013})}\BibitemShut {NoStop}%
\bibitem [{\citenamefont {Englert}\ and\ \citenamefont {W\'{o}dkiewicz}(2003)}]{Englert03}%
  \BibitemOpen
  \bibfield  {author} {\bibinfo {author} {\bibfnamefont {B.-G.}~\bibnamefont
  {Englert}}\ and\ \bibinfo {author} {\bibfnamefont {K.}\ \bibnamefont
  {W\'{o}dkiewicz}},\ }\bibfield  {title} {\bibinfo {title} {Tutorial notes on one-party and two-party Gaussian states},\ }\href
  {https://doi.org/10.1142/S0219749903000206} {\bibfield  {journal} {\bibinfo
  {journal} {Int. J. Quantum Inf.}\ }\textbf {\bibinfo {volume} {1}},\ \bibinfo
  {pages} {153} (\bibinfo {year} {2003}{\natexlab{b}})}\BibitemShut {NoStop}%
\bibitem [{\citenamefont {Tay}(2019)}]{Tay19b}%
  \BibitemOpen
  \bibfield  {author} {\bibinfo {author} {\bibfnamefont {B.}~\bibnamefont
  {Tay}},\ }\bibfield  {title} {\bibinfo {title} {Damping modes of harmonic
  oscillator in open quantum systems},\ }\href
  {https://doi.org/https://doi.org/10.1016/j.physa.2019.121119} {\bibfield
  {journal} {\bibinfo  {journal} {Physica A}\ }\textbf {\bibinfo {volume}
  {526}},\ \bibinfo {pages} {121119} (\bibinfo {year} {2019})}\BibitemShut
  {NoStop}%
\bibitem [{\citenamefont {Wu}\ and\ \citenamefont {Austin}(1977)}]{Wu77}%
  \BibitemOpen
  \bibfield  {author} {\bibinfo {author} {\bibfnamefont {T.}~\bibnamefont
  {Wu}}\ and\ \bibinfo {author} {\bibfnamefont {S.}~\bibnamefont {Austin}},\
  }\bibfield  {title} {\bibinfo {title} {Bose condensation in biosystems},\
  }\href {https://doi.org/https://doi.org/10.1016/0375-9601(77)90560-6}
  {\bibfield  {journal} {\bibinfo  {journal} {Phys. Lett. A}\ }\textbf
  {\bibinfo {volume} {64}},\ \bibinfo {pages} {151 } (\bibinfo {year}
  {1977})}\BibitemShut {NoStop}%
\bibitem [{\citenamefont {Kondepudi}\ and\ \citenamefont
  {Prigogine}(2014)}]{Kondepudi14}%
  \BibitemOpen
  \bibfield  {author} {\bibinfo {author} {\bibfnamefont {D.}~\bibnamefont
  {Kondepudi}}\ and\ \bibinfo {author} {\bibfnamefont {I.}~\bibnamefont
  {Prigogine}},\ }\href@noop {} {\emph {\bibinfo {title} {Modern
  Thermodynamics: From Heat Engines to Dissipative Structures}}},\ \bibinfo
  {edition} {2nd}\ ed.\ (\bibinfo  {publisher} {Wiley},\ \bibinfo {address}
  {UK},\ \bibinfo {year} {2014})\BibitemShut {NoStop}%
\end{thebibliography}
%apsrev4-2.bst 2019-01-14 (MD) hand-edited version of apsrev4-1.bst
%Control: key (0)
%Control: author (8) initials jnrlst
%Control: editor formatted (1) identically to author
%Control: production of article title (0) allowed
%Control: page (0) single
%Control: year (1) truncated
%Control: production of eprint (0) enabled
\providecommand{\noopsort}[1]{}\providecommand{\singleletter}[1]{#1}%

\end{document}